\documentclass[twocolumn, letter,nofootinbib,amsmath,amssymb,prl,superscriptaddress,tightenlines,preprintnumbers]{revtex4-1}
\usepackage[paperwidth=225mm,paperheight=297mm,centering,hmargin=2cm,vmargin=2.5cm]{geometry}
\usepackage{graphicx}
\usepackage{braket}
\usepackage[usenames]{color}
\usepackage{latexsym}
\usepackage{bm}
\usepackage{mathrsfs}
\usepackage[caption=false]{subfig}
\usepackage{slashed}

\usepackage{tikz}
\usetikzlibrary{arrows,decorations.pathmorphing,backgrounds,positioning,fit,petri,automata,shadows,calendar,mindmap,
decorations.markings,calc}

\usepackage[utf8]{inputenc}
\usepackage{graphicx}
\def\beq{\begin{equation}}
\def\eeq{\end{equation}}
\def\bea{\begin{eqnarray}}
\def\eea{\end{eqnarray}}

\newcommand{\eps}{\epsilon}

\def\p{\partial}

\def\sl{\slashed}

\def\calV{\mathcal{V}}
\def\calQ{\mathcal{Q}}
\def\calL{\mathcal{L}}
\def\calS{\mathcal{S}}

\newcommand{\semiloop}[4][]{%
\draw[#1] let \p1 = ($(#3)-(#2)$) in (#3) arc (#4:({#4+180}):({0.5*veclen(\x1,\y1)});)
}

\begin{document}

\newcount\hour \newcount\minute
\hour=\time \divide \hour by 60
\minute=\time
\count99=\hour \multiply \count99 by -60 \advance \minute by \count99
\newcommand{\mydate}{\ \today \ - \number\hour :00}

\tikzstyle{every picture}+=[remember picture]
\pgfdeclaredecoration{complete sines}{initial}
{
    \state{initial}[
        width=+0pt,
        next state=sine,
        persistent precomputation={\pgfmathsetmacro\matchinglength{
            \pgfdecoratedinputsegmentlength / int(\pgfdecoratedinputsegmentlength/\pgfdecorationsegmentlength)}
            \setlength{\pgfdecorationsegmentlength}{\matchinglength pt}
        }] {}
    \state{sine}[width=\pgfdecorationsegmentlength]{
        \pgfpathsine{\pgfpoint{0.25\pgfdecorationsegmentlength}{0.5\pgfdecorationsegmentamplitude}}
        \pgfpathcosine{\pgfpoint{0.25\pgfdecorationsegmentlength}{-0.5\pgfdecorationsegmentamplitude}}
        \pgfpathsine{\pgfpoint{0.25\pgfdecorationsegmentlength}{-0.5\pgfdecorationsegmentamplitude}}
        \pgfpathcosine{\pgfpoint{0.25\pgfdecorationsegmentlength}{0.5\pgfdecorationsegmentamplitude}}
}
    \state{final}{}
}
\tikzset{
fermion/.style={thick,draw=black, line cap=round, postaction={decorate},
    decoration={markings,mark=at position 0.6 with {\arrow[black]{latex}}}},
photon/.style={thick, line cap=round,decorate, draw=black,
    decoration={complete sines,amplitude=4pt, segment length=6pt}},
boson/.style={thick, line cap=round,decorate, draw=black,
    decoration={complete sines,amplitude=4pt,segment length=8pt}},
heavyboson/.style={ double, line cap=round,decorate, draw=black,
    decoration={complete sines,amplitude=4pt,segment length=7pt}},
gluon/.style={thick,line cap=round, decorate, draw=black,
    decoration={coil,aspect=1,amplitude=3pt, segment length=8pt}},
scalar/.style={dashed, thick,line cap=round, decorate, draw=black},
heavyscalar/.style={dashed, thick, double, line cap=round, decorate, draw=green},
heavyfermion/.style={thick, double, line cap=round, decorate, draw=blue},
ghost/.style={dotted, thick,line cap=round, decorate, draw=black},
->-/.style={decoration={
  markings,
  mark=at position 0.6 with {\arrow{>}}},postaction={decorate}}
 }

\makeatletter
\tikzset{
    position/.style args={#1 degrees from #2}{
        at=(#2.#1), anchor=#1+180, shift=(#1:\tikz@node@distance)
    }
}
\makeatother

\title{On the non-minimal character of the SMEFT}

\author{Yun~Jiang and Michael~Trott\\
{\it Niels Bohr International Academy,
University of Copenhagen, \\
Blegdamsvej 17, DK-2100 Copenhagen, Denmark}
}

\begin{abstract}
When integrating out unknown new physics sectors,
what is the minimal character of the Standard Model Effective Field Theory (SMEFT) that can result?
In this paper we focus on a particular aspect of this question:  ``How can one obtain only one dimension six operator in the SMEFT from a consistent tree level matching onto an unknown new physics sector?" We show why this requires conditions on the ultraviolet field content that do not indicate a stand alone ultraviolet complete scenario.
Further, we demonstrate how a dynamical origin of the ultraviolet scales assumed to exist in order to generate the masses of the heavy states integrated out generically induces more operators. Therefore,
our analysis indicates that the infrared limit captured from a new sector in consistent matchings induces multiple operators in the SMEFT quite generically.
Global data analyses in the SMEFT can and should accommodate this fact.  \end{abstract}

\maketitle
\newpage

\section{I. Introduction.}

Despite the null results on beyond the Standard Model (SM) resonance searches
from Run~I and II at LHC, the arguments in favor of physics beyond the SM are very strong. It is reasonable to expect
that the low energy effects of a new physics sector, which has a mass gap in its typical mass scale(s)\footnote{Schematically
denoted as $\Lambda$ and assumed to be in the $\sim \rm TeV$ range.} compared to the electroweak scale $v \simeq 246 \, {\rm GeV}$, could be resolved
in the future. This is particularly the case if $\Lambda \lesssim 4 \, \pi \, v$, which is consistent with expectations of ultraviolet (UV) physics motivated by naturalness concerns for the Higgs mass.
Broad classes of new physics scenarios consistent with this minimal decoupling assumption can be constrained efficiently
using effective field theory methods to analyse scattering data limited to energies $\sqrt{s} \sim v \ll \Lambda$.
This formalism has come to be known as
the Standard Model Effective Field Theory (SMEFT) recently \cite{Buchmuller:1985jz,Grzadkowski:2010es,Jenkins:2013zja,Jenkins:2013wua,Alonso:2013hga,Lehman:2014jma,Lehman:2015coa,Henning:2014wua,Henning:2015daa,Henning:2015alf}
where the SM is supplemented with a series of higher dimensional operators\footnote{The
complete sum of all non-redundant operators at each mass dimension order $d$ defines $\mathcal{L}_d$ here.}
\beq
\mathcal{L}_{\rm SMEFT} = \mathcal{L}_{\rm SM} + \mathcal{L}_{5}+ \mathcal{L}_{6} + \mathcal{L}_{7} + \mathcal{L}_{8} + \cdots
\eeq
In this approach, the null results of lower energy tests for physics beyond the SM can be consistent with beyond-the-SM UV physics in the $\sim \rm TeV$ mass scale range.
However, a large set of experimental measurements that test the symmetry breaking patterns of the SM must be accommodated.
This can be accomplished in a manner that avoids fine tuning.
In this work, we use
a symmetry assumption that new physics in the $\rm TeV$ mass range leads to a
$\mathcal{L}_6$ correction to the SM that (approximately) respects the
global symmetry group $\rm G = U(1)_B \otimes U(1)_L \otimes \rm SU(3)^5$, and in addition a discrete $\rm CP$
symmetry.\footnote{This hypothesis is not the only way to accommodate TeV field content, see for example
the discussion in Ref.~\cite{Giudice:2011ak}.}
Some of these symmetries cannot be exact, as the SM defines a minimal symmetry breaking in the SMEFT. However, the $\rm G$ and  $\rm CP$ symmetry breaking pattern of the SM can allow the effect of the UV physics sectors in the $\sim \rm TeV$ scales of experimental interest because of two reasons:  it follows a Minimal Flavor Violating (MFV) pattern for flavor changing measurements \cite{Chivukula:1987py,D'Ambrosio:2002ex,Cirigliano:2005ck}; and  it is proportional to the Jarlskog invariant \cite{Jarlskog:1985ht,Jarlskog:1985cw} in the case of SM $\rm CP$ violation.

In this paper, we consider matching patterns of operators that result from integrating out new physics sectors.
We study the effect of simultaneously integrating out not only the heavy fields, but also a UV sector
that generates the required heavy mass scale(s) $\Lambda > v$. We initially focus on the question of when, if ever, only one operator can be
obtained in such a tree level matching in our chosen basis for $\mathcal{L}_{6}$ \cite{Grzadkowski:2010es}. We examine this question in the SMEFT in terms of the matching effects of spin-\{1,1/2,0\} fields that can couple to the SM
through  ($d \leq 4$) mass dimension interactions. Higher spin composite fields (and spin towers) are possible and even required in the presence of UV confining strong interactions.
Similarly, when a UV sector with a strong interaction is present, there is no particular reason in general for tree level matchings, as opposed to non-perturbative matchings, to be the largest contribution to the Wilson coefficients. Such extra matching contributions are difficult to characterize (other than by using the full SMEFT formalism) and only reinforce our main point on the non-minimal character of the SMEFT,
so we do not focus on these contributions.

The number of operators induced in matching is operator basis dependent. However, the conditions uncovered on the UV field content to reduce the operator profile (i.e. the number of independent $\rm SU(3) \times SU(2) \times U(1)$ operators)
are still meaningful. The conditions can be framed in terms of symmetries and several simple observations on new physics spectra and dynamics that can generate a scale $\Lambda$, as we show. Practically speaking, most global analyses
are being constructed using the well defined Warsaw basis~\cite{Grzadkowski:2010es}, so we focus on this basis
when examining the one operator question. We use the notation $Q_i$ to denote an operator defined in the Warsaw basis in this work, and refer the
reader to Ref.~\cite{Grzadkowski:2010es} for the explicit operator definitions. Note also that we refer to one operator with the understanding that, consistent with our assumptions of $\rm G$ symmetry, flavor indices are not used to distinguish operators.

The structure of this paper is as follows.
Following a brief comment on the dimension-5 operator and Fermi theory in Section~II, we provide in Section ~III.A a comprehensive discussion on the SMEFT matching at tree level onto $\mathcal{L}_6$ when a massive spin-1 state present in a UV physics sector is integrated out. We focus this discussion on the ``one operator induced at tree level" question consistent with the assumed (approximate) $\rm G$ symmetry. We demonstrate why such a simple UV sector cannot be a complete scenario if a mechanism to generate the heavy state's mass is demanded. We then discuss the spin-1/2 case, drawing a similar conclusions in Section~III.B.  In Section~III.C we examine the case of integrating out a scalar field focused on the ``one operator"
question. We show how the scalar case is more subtle, but still argues for more operators when UV complete scenarios are demanded. Section~IV contains our conclusions.

\section{II. Two exceptional EFT cases}
\label{sec:dim5}

When considering the one operator question, we note that a few historical accidents in EFTs can be misleading.
First of all, $\mathcal{L}_5$ and $\mathcal{L}_d$ with $d\geq 6$ are distinct when considering this question.
Due to the charges of the SM field content, only one operator (with flavor indices) can be constructed in $\mathcal{L}_5$.
The operator that results \cite{Weinberg:1979sa,Wilczek:1979hc},
\beq
\mathcal{L}_5 = \frac{c_{ij}}{2} \, \left(\overline{L^c_{L,i}} \tilde{H}^\star\right) \left(\tilde{H}^\dagger \, L_{L,j}\right) + {\rm h.c.}
\eeq
is the well known example where one operator at a particular mass dimension does result when integrating out UV physics.\footnote{Here and below our notation with a $c$ superscript indicates a charge conjugate representation of a SM field.}
The interplay of global $\rm U(1)_L$ number violation and the constraints of the SM field's representations
leading to one operator in $\mathcal{L}_5$ is an exception that is not repeated at higher orders in the SMEFT operator expansion \cite{Lehman:2014jma,Lehman:2015coa,Henning:2014wua,Henning:2015daa,Henning:2015alf}.

Historically, Fermi theory has frequently been used as a prototypical EFT to build intuition.
This can be unfortunate, as Fermi theory is atypical and has a number of non-trivial accidental features that are not generic.
In Fermi theory, the four-fermion operator
\bea
Q_{\substack{\ell \ell}} = \left(\overline{L_L} \, \gamma^\mu  L_L \right) \left(\overline{L_L} \, \gamma_\mu L_L\right),
\eea
is generated when the $W$ boson is integrated out. This effective operator is used in the process $\mu^- \rightarrow e^- + \bar{\nu}_e+ \nu_\mu$
to infer the Fermi constant, $G_F$.
The UV sector in the case of Fermi theory is the SM which does induce a series of other operators at tree level, in addition to the operator $Q_{\ell \ell}$. These
 four-fermion operators are due to the Higgs field and the $Z$ boson. However, the highly suppressed Yukawa couplings of the SM Higgs to light fermions leads to an exceptional situation numerically in terms of the operator profiles.
The small Yukawa couplings are not formally the consequence of a fine tuning, as they are protected by the full chiral symmetry of the SM.
More discussion on the accidents in Fermi theory, and how it is commonly misunderstood, can be found in Ref.~\cite{Gavela:2016bzc}.

Arguably, there is some theoretical evidence based on the structure and particle content of the SM in the direction of embedding this model into $\rm SU(5)$ or $\rm SU(10)$, see for example the arguments in Ref.~\cite{Georgi:1974sy}. This could be interpreted as a hint to an underlying theory, similar to the chiral structure of the SM being a low energy hint of its UV structure. However, the problems of $\rm TeV$ scale grand unified theories are very well known. In this work we make a more phenomenologically motivated choice
and assume approximate $\rm G$ symmetry (and $\rm CP$ symmetry).

\section{III. $G$ symmetric tree level matchings}
\subsection{A. Spin $1$ states}

\begin{table}[t]
\begin{center}
\begin{tabular}{|c|c|c|c|c|c|c|}
  \hline
   Case &$\rm SU(3)_C$ & $\rm SU(2)_{L}$ &$\rm U(1)_Y$  &$\rm G_Q$ &$\rm G_L$ & Couples to \\
  \hline
$\calV_{\rm I}^{(1,8)}$ &  1,8 & 1& 0  & (1,1,1) &  (1,1) & ${\bar{d}_R} \, \gamma^\mu \, d_R$ \\[2pt]
$\calV_{\rm II}^{(1,8)}$ &  1,8 & 1& 0  & (1,1,1) & (1,1) &  ${\bar{u}_R} \, \gamma^\mu \, u_R$ \\[2pt]
$\calV_{\rm III}^{(1,8)}$ &  1,8 & 1& 0  & (1,1,1) & (1,1) &  ${\bar{Q}_L} \, \gamma^\mu \, Q_L$ \\[2pt]
$\calV_{\rm IV}^{(1,8)}$ &  1,8 & 3& 0  & (1,1,1) & (1,1) &  ${\bar{Q}_L} \sigma^I\, \gamma^\mu \, Q_L$ \\[2pt]
  \hline
  \hline
$\calV_{\rm V}^{(1,8)}$ &  1,8 & 1& 0  & (1,8,1) & (1,1) &  ${\bar{d}_R} \, \gamma^\mu \, d_R$ \\[2pt]
$\calV_{\rm VI}^{(1,8)}$ &  1,8 & 1& 0  & (8,1,1) & (1,1) &  ${\bar{u}_R} \, \gamma^\mu \, u_R$ \\[2pt]
$\calV_{\rm VII}^{(1,8)}$ &   1,8 & 1& -1  & ($\bar{3}$,3,1) & (1,1) &  ${\bar{d}_R} \, \gamma^\mu \, u_R$ \\[2pt]
$\calV_{\rm VIII}^{(1,8)}$ &  1,8 & 1& 0  & (1,1,8) & (1,1) &  ${\bar{Q}_L} \, \gamma^\mu \, Q_L$ \\[2pt]
$\calV_{\rm IX}^{(1,8)}$ &  1,8 & 3& 0  & (1,1,8) & (1,1) &  ${\bar{Q}_L} \sigma^I \, \gamma^\mu \, Q_L$ \\[2pt]
$\calV_{\rm X}^{(\bar{3},6)}$ &  $\bar{3}$,6 & 2& -1/6  & (1,3,3) & (1,1) &  ${\bar{d}_R} \, \gamma^\mu \, Q_L^c$ \\[2pt]
$\calV_{\rm XI}^{(\bar{3},6)}$ & $\bar{3}$,6 & 2& 5/6  & (3,1,3) & (1,1) &  ${\bar{u}_R} \, \gamma^\mu \, Q_L^c$ \\[2pt]
  \hline
  \end{tabular}
\end{center}
\caption{\label{quarkvectors} Vector representations \cite{Grinstein:2011yv,Grinstein:2011dz} consistent with our assumptions.
The first three rows are the same field sub-classified.
Superscripts on the field label indicate the representation under color. The Gell-Mann matrix $T^A$ (for both color and flavor $8$'s)
is present but suppressed in the coupling to some fermion bi-linears. $\sigma_I$ is the Pauli matrix. The table largely follows from the $\rm SU(3)$ group relations $\bf 3 \otimes \bar{\bf 3} = \bf 1 \oplus \bf 8$ and  $ \bf 3  \otimes \bf 3 = \bf 6 \oplus \bar{\bf 3}$.}
\end{table}

Spin-1 fields that couple to the SM quark bi-linears in the manner assumed are given by Table~\ref{quarkvectors} \cite{Grinstein:2011yv,Grinstein:2011dz,delAguila:2010mx}.
The requirement of linear couplings of mass dimension less than four, together with Lorentz symmetry and invariance under the full SM gauge group constrains the possible quantum numbers of UV field content.  Fields with other
representations that give SMEFT matchings respecting $\rm G$ are possible, if these conditions are relaxed.
Our notation is that $Q^c$, $L^c$ are the right handed conjugate doublet fields of the SM fermions.
The global flavor symmetry in the quark and lepton sectors are defined as
\bea
\rm G_Q &=& \rm SU(3)_{u_R} \times SU(3)_{d_R} \times SU(3)_{Q_L}, \\
\rm G_L &=& \rm SU(3)_{L_L}  \times SU(3)_{e_R}.
\eea
It is also possible to have a vector field couple to lepton bi-linears, to quark-lepton bi-linears or have an interaction with the SM Higgs field.
We list the corresponding fields in Table~\ref{leptonvectors} and Table~\ref{scalarvectors}.
Cases $\calV_{\rm XII}$, $\calV_{\rm XIII}$ have fields that carry a global lepton number and $\calV_{\rm XIV} - \calV_{\rm XX}$ carry both lepton and baryon numbers.
Although counterintuitive, UV fields that carry flavor quantum numbers do not necessarily lead to lower energy signatures
of flavor violation -- outside of the MFV pattern. Similarly, fields carrying lepton number do not necessarily lead to lower energy signatures
of lepton flavor violation at tree level.\footnote{This was previously noted in Ref.~\cite{delAguila:2010mx} in the lepton number case for flavor singlet fields.} Fields that carry both lepton and baryon number are potentially more problematic in inducing proton decay, but such phenomenological constraints are not the focus of this paper.
\begin{table}[t]
\begin{center}
\begin{tabular}[t]{|c|c|c|c|c|c|c|}
  \hline
   Case &$\rm SU(3)_C$ & $\rm SU(2)_{L}$ &$\rm U(1)_Y$  &$\rm G_Q$ &$\rm G_L$ & Couples to \\
  \hline
      $\calV_{\rm I}^{(1)}$ &  1 & 1& 0  & (1,1,1) & (1,1) &  ${\bar{e}_R} \, \gamma^\mu \, e_R$ \\[2pt]
    $\calV_{\rm I}^{(1)}$ &  1 & 1& 0  & (1,1,1) & (1,1) &  ${\bar{L}_L} \, \gamma^\mu \, L_L$ \\[2pt]
  $\calV_{\rm IV}^{(1)}$ &  1 & 3& 0  & (1,1,1) & (1,1) &  ${\bar{L}_L} \sigma^I\, \gamma^\mu \, L_L$ \\[2pt]
  \hline
  \hline
$\calV_{\rm XII}$ &  1 & 2 & 3/2  & (1,1,1) &  ($\bar{3}$,$\bar{3}$) & ${\bar{L}^c_L} \, \gamma^\mu \, e_R$ \\[2pt]
$\calV_{\rm XIII}$ & 1 & 1 & 0 & (1,1,1) & (1,8) &  ${\bar{e}_R} \, \gamma^\mu \, e_R$ \\[2pt]
$\calV_{\rm XIV}$ & $\bar{3}$ & 2 & -1/6  & ($\bar{3}$,1,1) &  ($\bar{3}$,1) & ${\bar{L}^c_L} \, \gamma^\mu \, u_R$ \\[2pt]
$\calV_{\rm XV}$ & $\bar{3}$ & 2 & 5/6  & (1,$\bar{3}$,1) &  ($\bar{3}$,1) & ${\bar{L}^c_L} \, \gamma^\mu \, d_R$ \\[2pt]
$\calV_{\rm XVI}$ & $\bar{3}$ & 1 & -2/3  & (1,$\bar{3}$,1) & (1,3) &  ${\bar{e}_R} \, \gamma^\mu \, d_R$ \\[2pt]
$\calV_{\rm XVII}$ & $\bar{3}$ & 1 & -5/3  & ($\bar{3}$,1,1) & (1,3) &  ${\bar{e}_R} \, \gamma^\mu \, u_R$ \\[2pt]
$\calV_{\rm XVIII}$ & $3$ & 2 & -5/6  & (1,1,3) & (1,3) &  ${\bar{e}_R} \, \gamma^\mu \, Q^c_L$ \\[2pt]
$\calV_{\rm XIX}$ & $\bar{3}$ & 1 & -2/3  & (1,1,$\bar{3}$) & (3,1) &  ${\bar{L}_L} \, \gamma^\mu \, Q_L$ \\[2pt]
$\calV_{\rm XX}$ & $\bar{3}$ & 3 & -2/3  & (1,1,$\bar{3}$) & (3,1) &  ${\bar{L}_L} \sigma^I\, \gamma^\mu \, Q_L$ \\[2pt]
$\calV_{\rm XXI}$ & 1 & 1 & 0  & (1,1,1) & (8,1) &  ${\bar{L}_L} \gamma^\mu \, {\bar{L}_L}$ \\[2pt]
$\calV_{\rm XXII}$ & 1 & 3 & 0  & (1,1,1) & (8,1) &  ${\bar{L}_L} \sigma^I \, \gamma^\mu \, {\bar{L}_L}$ \\[2pt]
  \hline
  \end{tabular}
\end{center}
\caption{\label{leptonvectors}Different vector representations that couple to fermion bi-linears respecting $\rm G$, without the insertion of a Yukawa matrix.}
\begin{center}
\begin{tabular}[t]{|c|c|c|c|c|c|c|}
  \hline
   Case &$\rm SU(3)_C$ & $\rm SU(2)_{L}$ &$\rm U(1)_Y$  &$\rm G_Q$ &$\rm G_L$ & Couples to \\
  \hline
$\calV_{\rm I}^{(1)} \! \! ,\calV_{\rm II}^{(1)} \! \! ,\calV_{\rm III}^{(1)}$ &  1 & 1& 0  & (1,1,1) &  (1,1) & $H^\dagger i D^\mu H$ \\[2pt]
$\calV_{\rm IV}^{(1)}$ &  1 & 3& 0  & (1,1,1) & (1,1) &  $H^\dagger \sigma^I i D^\mu H$ \\[2pt]
  \hline
  \hline
$\calV_{\rm XXIII}^{(1)}$ &  1 & 1& -1  & (1,1,1) & (1,1) &  $H^T i D_\mu H$ \\[2pt]
$\calV_{\rm XXIV}^{(1)}$ &  1 & 3& -1  & (1,1,1) & (1,1) &  $H^T i \sigma_I D_\mu H$ \\[2pt]
  \hline
  \end{tabular}
\end{center}
\caption{\label{scalarvectors} Vector representations coupling to currents constructed from Higgs fields.}
\end{table}

\vspace*{-0.5mm}
\subsubsection{Dimension-6 operator matching}
\vspace*{-1mm}
Solving the classical equations of motion (EOM) for the heavy vector fields and substituting the classical solution into the Lagrangian
results in a direct tree level matching in terms of a product of currents. We define the currents as
\beq
J_a =\{J_\psi^\mu,J_H^\mu \} = \{\bar{\psi} \, \gamma_\mu \otimes \psi,(D_\mu H)^\dagger \otimes \Phi\},
\eeq
and the tree level matching is given by
\beq
\Delta \mathcal{L}_6 \supset - \frac{1}{M_V^2} \,
(J_a^\mu)^\dagger J_b^{\mu}.
\label{oureffective}
\eeq
Here $\Phi$ represents $H$ or $\tilde{H} = i \, \sigma_2 \, H^\star$ and  $\otimes$ indicates a group product characterized by the $\rm SU(2)_L$ representation of vector fields.\footnote{This notation is consistent with Ref.~\cite{delAguila:2010mx}. Note also that a further current of the form $D^\mu F_{\mu \nu}$ with $F = \{B,W,G\}$ is redundant \cite{delAguila:2010mx}.}
For a vector field of the form considered in Tables \ref{quarkvectors},~\ref{leptonvectors} and \ref{scalarvectors}, the current product falls into one of three types:
\begin{itemize}
\item{four-fermion: $(J_\psi^\mu)^\dagger \, J_{\psi,\mu}$,}
\item{scalar derivative: $(J_H^\mu)^\dagger \, J_{H,\mu}$,}
\item{mixed scalar-fermion: $(J_\psi^\mu)^\dagger \, J_{H,\mu}, (J_H^\mu)^\dagger \, J_{\psi,\mu}$.}
\end{itemize}
We have systematically examined the profile in terms of operators obtained in tree level matchings to the Warsaw basis from the fields listed in Tables~\ref{quarkvectors},\ref{leptonvectors} and \ref{scalarvectors}, finding the following rule:

\vspace{1.5mm}
{\it Flavour singlet vector fields that do not break $\rm G_Q \times G_L$ induce more than one operator at tree level when
matching onto the SMEFT Warsaw basis.}

\begin{center}
\begin{table}[t!]
\hspace*{-5mm}
\begin{tabular}{|c||c|c|}
  \hline
   Case & $Q_i$ generated at tree level   \\
  \hline
$\calV_{\rm IV}^{(1)}$ & $Q_{ll}, Q^{(3)}_{qq,lq}, Q^{(3)}_{Hq,Hl}, Q_H, Q_{H \rm D}, Q_{H \Box}, Q_{eH}, Q_{uH}, Q_{dH}$   \\[2pt]
  \hline
  \hline
$\calV_{\rm IV}^{(8)}$ & $ Q^{(1)}_{qq}, Q^{(3)}_{qq}$ \\[2pt]
$\calV_{\rm IX}^{(8)}$ &   $ Q^{(1)}_{qq}, Q^{(3)}_{qq}$ \\[2pt]
$\calV_{\rm XX}$ &  $Q^{(1)}_{lq}, Q^{(3)}_{lq}$  \\
\hline
\hline
$\calV_{\rm XXIII}^{(1)}$ &  $Q_H, Q_{H \rm D}, Q_{H \Box}, Q_{eH}, Q_{uH}, Q_{dH}$ \\[2pt]
$\calV_{\rm XXIV}^{(1)}$ &  $Q_H, Q_{H \rm D}, Q_{H \Box}, Q_{eH}, Q_{uH}, Q_{dH}$  \\[2pt]
  \hline
  \end{tabular}
  \vspace*{1mm}
\caption{\label{vectorsmanyops} Examples of the sets of $\mathcal{L}_6$ operators in the SMEFT obtained by integrating out various massive vectors.}
\end{table}
\end{center}
\begin{center}


\begin{table}[t!]
\begin{tabular}{|c||c|c|c|c|}
  \hline
   Case & Op  & $\rm U(1)_Y$ & $G_Q,G_L$ Spurion  \\
  \hline
$\calV_{\rm VIII}^{(1)}$ &  $Q^{(1)}_{qq}$ & 0 & $T^A \, Y_u^\dagger Y_u, T^A \, Y_d^\dagger Y_d$   \\[2pt]
$\calV_{\rm IX}^{(1)}$ &  $Q^{(1)}_{qq}$ & 0 & $T^A \, Y_u^\dagger Y_u, T^A \, Y_d^\dagger Y_d$  \\[2pt]
$\calV_{\rm XIX}$ &  $Q^{(1)}_{lq}$ & -2/3 & /  \\
\hline
\hline
$\calV_{\rm X}^{(\bar{3},6)}$ &  $Q^{(1)}_{qd}$ & -1/6 &/  \\[2pt]
$\calV_{\rm XI}^{(\bar{3},6)}$ &  $Q^{(1)}_{qu}$ &5/6  &/ \\[2pt]
$\calV_{\rm XVIII}$ &  $Q_{qe}$ & -5/6 &/  \\
$\calV_{\rm XII}$ &  $Q_{le}$ & 3/2 &/  \\
$\calV_{\rm XIV}$ &  $Q_{lu}$ & -1/6 &/  \\
$\calV_{\rm XV}$ &  $Q_{ld}$ & 5/6  &/  \\
  \hline
  \hline
$\calV_{\rm V}^{(1)}$ & $Q_{dd}$ & 0 &$T^A \, Y_d^\dagger Y_d$  \\[2pt]
$\calV_{\rm VI}^{(1)}$ & $Q_{uu}$ & 0 &$T^A \, Y_u^\dagger Y_u$  \\[2pt]
$\calV_{\rm VII}^{(1)}$ & $Q^{(1)}_{ud}$ & -1 &$Y_d^\dagger Y_u$   \\[2pt]
$\calV_{\rm XIII}$ &  $Q_{ee}$ & 0 &$T^A \, Y_e^\dagger Y_e$  \\
$\calV_{\rm XVI}$ &  $Q_{ed}$ & -2/3 &/  \\
$\calV_{\rm XVII}$ &  $Q_{eu}$ & -5/3 &/  \\
  \hline
  \end{tabular}
\vspace*{1mm}
\caption{Operators induced at tree level when the massive vector case is integrated out.
The cases are grouped in the table into the chiral $(J_\psi^\mu)^\dagger \, J_{\psi,\mu}$ operator classes induced.
The top section refers to $\rm LLLL$ operators. The middle section of the table refers to $\rm LLRR$ operators.
The bottom section of the table refers to $\rm RRRR$ operators induced at tree level.}
\label{vectors1}
\end{table}
\end{center}
This result is easy to demonstrate. Fields that are $\rm SU(3)_C$ and $\rm SU(2)_L$ singlets couple to (quark and lepton) fermion fields
and also the scalar currents, inducing a large number of operators at tree level. A vector field can be made
to couple to the left-handed doublets by assigning the field to a $\bf 3$ of $\rm SU(2)_L$. The scalar and leptonic couplings
can be removed by assigning the field to a $\bf 8$ of $\rm SU(3)_C$. In this case the operator profile is reduced to at least
two $(J_\psi^\mu)^\dagger \, J_{\psi,\mu}$ operators via the relation \cite{Grzadkowski:2010es}
\begin{align}
({\bar{Q}_L^p} \sigma^I  \, T^A\, \gamma^\mu \, Q_L^r)({\bar{Q}_L^s} \sigma^I  \, T^A\, \gamma^\mu \, Q_L^t) = \nonumber \\
 - \frac{1}{4} Q_{\substack{qq \\ ptsr}}^{(3)}
+ \frac{3}{4} Q_{\substack{qq \\ ptsr}}^{(1)} - \frac{1}{6} Q_{\substack{qq \\ prst}}^{(3)},
\end{align}
here $p,r,s,t$ are flavor indices. We show some examples of the multiple operators induced when integrating out
vector fields at tree level in Table \ref{vectorsmanyops}.
Introducing $\rm G_Q \times G_L$ symmetry, vector fields can
be reduced in their infrared (IR) SMEFT operator profile to one operator in the Warsaw basis
in the limit of vanishing Yukawa matrices; see Table \ref{vectors1}.
Note that with the exception of case $\calV_{\rm VII}^{(1)}$ which has a
bi-linear flavor breaking spurion in $Y_d^\dagger$ and $Y_u$, the presence of a
$\rm U(1)_Y$ charge is also associated with the lack of Higgs scalar currents induced. This has an important consequence when
the self interactions of the vector are studied for unitarity violation, as will be discussed shortly.

A spurion analysis allows the corrections due to
the nonzero Yukawa matrices of the SM (that break the flavor symmetry in a phenomenologically safe MFV pattern) to be systematically studied. We define the SM Yukawa matrices $Y_u, Y_d, Y_e$ as
\begin{align}
\mathcal{L}_{Y} &= -(Y_u)^p_r \, \bar{u}_{R,p} \, Q_L^r \, \tilde{H}^\dagger - (Y_d)^p_r \, \bar{d}_{R,p} \, Q_L^r \, H^\dagger \nonumber \\
&\quad \, -  (Y_e)^p_r \, \bar{e}_{R,p} \, L_L^r \, H^\dagger + {\rm h.c.}
\end{align}
$\rm G_Q \times G_L$ symmetry is restored if we endow the Yukawa matrices with the transformation properties under $\rm \{G_Q,G_L\}$
\bea
Y_u &\sim& (3,1,\bar{3},1,1), \quad Y_d \sim (1,3,\bar{3},1,1), \nonumber \\
Y_e &\sim& (1,1,1,\bar{3},3).
\eea
Introducing $\rm G_Q \times G_L$ symmetry breaking when the $Y_u, Y_d, Y_e$ matrices take on their SM values
gives more operators at tree level for fields with flavor quantum numbers. On general grounds, the $(J_H^\mu)^\dagger \, J_{H,\mu}$ current products are induced proportional to two spurions breaking the flavor symmetry, and the $(J_\psi^\mu)^\dagger \, J_{H,\mu}, (J_H^\mu)^\dagger \, J_{\psi,\mu}$ current products
are induced proportional to one flavor breaking spurion insertion. Here we refer to the spurions listed in Table \ref{vectors1} that are bi-linear in Yukawa matrices. As a specific example consider $\calV_{\rm VIII}^{(1)}$ that is
a $\bf 8$ under $\rm SU(3)_{Q_L}$. The Lagrangian\footnote{Recall the flavor adjoint $\bf 8$ representation is real.} is given by $\mathcal{L}_{\rm SM} +\mathcal{L}_{\calV_{\rm VIII}^{(1)}}$ where
\begin{align}
\label{sampleV}
\mathcal{L}_{\calV_{\rm VIII}^{(1)}} &= -\frac{1}{2} \left(D_\mu \calV_\nu \, D^\mu \calV^\nu - D_\mu \calV_\nu \, D^\nu \calV^\mu \right) -  \frac{M^2_{\calV}}{2}
\calV_\nu \calV^\nu \nonumber  \\
& \, \quad + \left(\lambda_\calV \calV_{\mu,A} T^A \, Y_u^\dagger Y_u \, (D^\mu H)^\dagger H + {\rm h.c.} \right), \\
&\, \quad +  g_\calV \calV_{\mu,A} ({\bar{Q}_L}  T^A \gamma^\mu  Q_L) + \cdots. \nonumber
\end{align}
Note that the largest spurion that restores the flavor symmetry for the second line is $T^A Y_u^\dagger Y_u$ and some indices are suppressed
in Eqn.~(\ref{sampleV}). The additional spurion breaking proportional to $Y_d^\dagger Y_d$ is neglected in what follows.
Integrating by parts and the EOM for the vector field
are used to manipulate the derivative to appear as shown on the second line in Eqn.~(\ref{sampleV}).
Integrating out the field $\calV_{\rm VIII}^{(1)}$ using the classical EOM gives
\bea
\Delta \mathcal{L}_6 &\supset& \mathcal{}\frac{g_\calV^2}{4 \, M^2_{\calV}} \left[Q_{\substack{qq \\ rs sr}}^{(1)} - \frac{1}{3} \,Q_{\substack{qq \\ rr ss}}^{(1)} \right]  \nonumber\\
&+& \frac{1}{4 \, M^2_{\calV}} \,  \Big[ (({\rm Im}\lambda_\calV)^2 -({\rm Re}\lambda_\calV)^2) Q_{H\Box} + 4 ({\rm Im}\lambda_\calV)^2 \, Q_{HD}  \nonumber\\
&&\quad\quad\quad  + 2i ({\rm Re}\lambda_\calV) ({\rm Im}\lambda_\calV) (Y_b^\dagger Q_{bH} - Y_b Q^\dagger_{bH}) \nonumber\\
&&\quad\quad\quad - 2i ({\rm Re}\lambda_\calV) ({\rm Im}\lambda_\calV) (Y_u^\dagger Q_{uH} - Y_u Q^\dagger_{uH})
 \Big] \nonumber \\
& & \quad \quad \times \left[{\rm Tr}[(Y_u^\dagger Y_u)(Y_u^\dagger Y_u)] - \frac{({\rm diag}(Y_u^\dagger Y_u))^2}{3}\right]  \\
&-& \frac{g_\calV{\rm Im}[\lambda_\calV]}{2 M^2_{\calV}} Q_{\substack{Hq \\ pr}}^{(1)} \left[(Y_u^\dagger Y_u)^p_r - \frac{{\rm diag}(Y_u^\dagger Y_u)}{3} \delta_{pr} \right] \nonumber \\
&+& i \, \frac{g_\calV{\rm Re}[\lambda_\calV]}{2 M^2_{\calV}} \left[ ((Y_u^\dagger Y_u) Y^\dagger_a )^m_i   Q_{\substack{aH \\ im}} -  (Y_a (Y_u^\dagger Y_u))_m^i  Q^\dagger_{\substack{aH \\ mi}} \right]\nonumber \\
&-& i \, \frac{g_\calV  {\rm Re}[\lambda_\calV]}{6 M^2_{\calV}} {\rm Tr}[Y_u^\dagger Y_u] \left[(Y_a^\dagger)^m_i Q_{\substack{aH \\ im}} - (Y_a)^i_m Q^\dagger_{\substack{aH \\ mi}}\right]
 , \nonumber
\eea
where the dummy labels $a$ and $b$ are summed over $\{u,d\}$ and $\{e, d\}$, respectively. A similar pattern of matchings onto the class 3 ($D^2 H^4$), 5 ($H^3 \bar{\psi} \psi$) and 7 ($H^2 D \bar{\psi} \psi$) operators
of the Warsaw basis is present for almost all color singlet fields with flavor quantum numbers listed in Tables~\ref{leptonvectors} and~\ref{scalarvectors}.
The exceptional case is the field $\calV_{\rm XII}$ whose non-trivial $\rm SU(2)_L$ representation and $\rm U(1)_Y$ charge forbids a scalar current from being induced
at tree level in this manner.

The pattern of tree level matchings is strongly dictated by the charges and representations of the UV fields under $\rm SU(3)_C \times SU(2)_L \times U(1)_Y$,
 $\rm G_Q$ and $\rm G_L$. We emphasize, data fits to subsets of operators in the SMEFT formalism can be justified by appealing to
UV field content with $\rm U(1)_Y$ charges and non-trivial representations under SM groups
when only retaining tree level matching contributions. See Table \ref{vectors1} for details on cases
that generate only one operator at a time.

This conclusion is subject to the following qualifications. First, the single operators obtained in tree level matchings to the
vectors in Tables~\ref{leptonvectors},~\ref{scalarvectors} are limited to $(J_\psi^\mu)^\dagger \, J_{\psi,\mu}$ operator forms. Such operators
at LHC contribute to continuum parton production in a fashion dictated by the power counting of the theory. Conversely, the precise measurements
made on a scattering through a SM resonance (with mass $M$ and width $\Gamma$) parametrically has a $\Gamma/M$ suppression, compared to the leading
resonant behavior, when considering the interference with
$(J_\psi^\mu)^\dagger \, J_{\psi,\mu}$ operators.

Second, as $y_t \simeq 1$, a
flavor symmetry spurion breaking proportional to only powers of $Y_u$ can induce operators of
class 3, 5 and 7 without significant numerical suppression. This makes it difficult to justify ``one at a time"
data fits to $(J_\psi^\mu)^\dagger \, J_{\psi,\mu}$ SMEFT operators with up quark field content (consistent with our assumptions). On the other hand,
one at a time data fits to $(J_\psi^\mu)^\dagger \, J_{\psi,\mu}$ operators that only have leptonic or down quark field content can be potentially justified.
In these cases the induced scalar currents proportional to MFV like
flavor breaking spurious are numerically suppressed compared to pure up quark spurions by at least $y_b/y_t \sim 10^{-2}$.

Finally, we also note that we never obtain only one operator in such a tree level matching that involves the Higgs field,
in the cases of massive vector UV field content considered.

\vspace*{-2mm}
\subsubsection{Arguments against orphaned vectors.}
\label{nouglyorphans}
\vspace*{-8mm}

\begin{center}
\begin{table}[t!]
\begin{tabular}{|c||c|c|c|c|c|c|}
  \hline
   Case & $\frac{16 \pi^2 \epsilon \, \delta_{Z_3}}{g_{\calV}^2 \langle \mathcal{O} \rangle}$  & $\frac{-16 \pi^2 \epsilon \, \delta_{Z_{\bar{\psi}}}}{g_{\calV}^2}$ & $\frac{-16 \pi^2 \epsilon \, \delta_{Z_\psi}}{g_{\calV}^2}$ & $\frac{-16 \pi^2 \epsilon \, \delta_{Z_V}}{g_{\calV}^2}$ & $\beta_y$ \\
  \hline
$\calV_{\rm VIII}^{(1)}$ & $\mathcal{F}(3_{\rm Fl})$  & $C_F^{(3_{\rm Fl})}$ & $C_F^{(3_{\rm Fl})}$ & ${4\over 3} \left({1\over 2}\right)_{\rm Fl} \!\cdot 3_{\rm C}$  & $+$ \\[2pt]
$\calV_{\rm IX}^{(1)}$ & $\mathcal{F}(2) \mathcal{F}(3_{\rm Fl}) $  & $C_F^{(2)} C_F^{(3_{\rm Fl})}$ & $C_F^{(2)} C_F^{(3_{\rm Fl})}$ & ${2 \cdot 3_{\rm C} \over 3} \left({1\over 2}\right)_{\rm Fl} \! \! \left({1\over 2}\right)_{\rm L} $  & $+$ \\[2pt]
$\calV_{\rm XIX}$ & 1 & $ 3_{\rm Fl} \! \cdot \! 3_{\rm C}$  & $3_{\rm Fl}$ & ${2\over 3} \cdot 2$ & $+$ \\
\hline
\hline
$\calV_{\rm X,XI}^{(\bar{3})}$ & $-1_{\rm C}$ & $3_{\rm Fl} (-2)_{\rm C} \! \cdot \!2$ & $3_{\rm Fl}(-2)_{\rm C}$ & ${2\over 3} \!\cdot\! (-1)_{\rm C} $ &  $-$ \\[2pt]
$\calV_{\rm X,XI}^{(6)}$ & $ 1$ & $3_{\rm Fl}\!\cdot \!1_{\rm C}\!\cdot \!2$ & $3_{\rm Fl}\!\cdot \!1_{\rm C}$ & ${2\over 3} \cdot\! 1_{\rm C}$ &  $+$ \\[2pt]
$\calV_{\rm XVIII}$  & 1 & $ 3_{\rm Fl} \!\cdot \! 3_{\rm C}\!\cdot \! 2$ &  $3_{\rm Fl}$ & ${2\over 3} $ &  $+$ \\
$\calV_{\rm XII}$ & 1 &  $3_{\rm Fl}$ & $3_{\rm Fl}\!\cdot \! 2$ & ${2\over 3}$ &  $+$ \\
$\calV_{\rm XIV}$ & 1 &  $3_{\rm Fl} \!\cdot \! 3_{\rm C}$ & $3_{\rm Fl}\!\cdot \! 2$ & ${2\over 3}$ &  $+$ \\
$\calV_{\rm XV}$ & 1 &  $3_{\rm Fl} \!\cdot \! 3_{\rm C}$ & $3_{\rm Fl}\!\cdot \! 2$ & ${2\over 3} $ &  $+$  \\
  \hline
  \hline
$\calV_{\rm V}^{(1)}$ & $\mathcal{F}(3_{\rm Fl}) $  & $C_F^{(3_{\rm Fl})}$ & $C_F^{(3_{\rm Fl})}$ & ${2\over 3} \left({1\over 2}\right)_{\rm Fl} \!\cdot \! 3_{\rm C} $ & $+$ \\[2pt]
$\calV_{\rm VI}^{(1)}$ & $\mathcal{F}(3_{\rm Fl}) $  & $C_F^{(3_{\rm Fl})}$ & $C_F^{(3_{\rm Fl})}$ & ${2\over 3} \left({1\over 2}\right)_{\rm Fl} \!\cdot \! 3_{\rm C} $  & $+$ \\[2pt]
$\calV_{\rm VII}^{(1)}$ & 1  & $3_{\rm Fl}$ & $3_{\rm Fl}$ & ${2\over 3} \!\cdot \! 3_{\rm C}$  & $ + $  \\[2pt]
$\calV_{\rm XIII}$ & $\mathcal{F}(3_{\rm Fl})$ & $C_F^{(3_{\rm Fl})}$ & $C_F^{(3_{\rm Fl})}$ & ${2\over 3} \left({1\over 2}\right)_{\rm Fl}  $  & $+$ \\
$\calV_{\rm XVI}$ & 1 & $3_{\rm Fl} \!\cdot \! 3_{\rm C}$ & $3_{\rm Fl}$ & ${2\over 3} $  & $ + $ \\
$\calV_{\rm XVII}$ & 1 &  $3_{\rm Fl} \!\cdot \! 3_{\rm C}$ & $3_{\rm Fl}$ & ${2\over 3} $ & $+ $ \\
  \hline
  \end{tabular}
\caption{One loop renormalization results. Here $\langle O \rangle$ indicates the
matrix element of the vector-fermion bilinear interaction term and $\delta_{Z_3}$ corresponds to the divergence present in this three point interaction
at one loop from the vector-fermion coupling. The notation is such that $\mathcal{F}(N)\equiv C_F^{(N)}-{1 \over 2} N$ with $C_F^{(N)}={N^2-1 \over 2N}$. We have labeled several of
the numerical factors in the table with the group space ($\rm SU(3)_C$, $\rm SU(3)_{Fl}$, $\rm SU(2)_L$) that generates them, with the subscript $Fl$
indicating a $\rm SU(3)$ flavour group.}
\label{int-vectorRGE}
\end{table}
\end{center}

The vector fields listed in Tables~\ref{quarkvectors}, \ref{leptonvectors} and \ref{scalarvectors} inducing a single $\mathcal{L}_6$ SMEFT operator
at tree level, carry at least one non-trivial representation under the SM gauge symmetry and flavor symmetries.\footnote{In all cases but one,
multiple non-trivial representations are present.
The one exceptional case is $\calV_{\rm XIII}$ which is only an $\bf 8$ under $\rm SU(3)_{e_R}$.}
Non-trivial representations and $\rm U(1)_Y$ charges reduce the interactions for SM particles with the new sector,
which consequently minimizes the IR SMEFT operator profile. However, such fields in general do not indicate
a stand alone UV complete scenario (where the vector could be an ``orphan") for the following reasons.

\begin{figure}[t!]
\centering
\begin{tikzpicture}[node distance=0.7cm, rounded corners=0pt, line cap=round]
\begin{scope}[]
\coordinate[] (v1);
\coordinate[left=of v1] (a1);
\coordinate[right=1.0cm of v1] (v2);
\coordinate[right=of v2] (b1);
\draw[heavyboson] (a1)node[left]{$\mathcal{V}$} -- (v1);
\draw[heavyboson](v2)--(b1)node[right]{$\mathcal{V}$};
\semiloop[fermion]{v1}{v2}{0};
\semiloop[fermion]{v2}{v1}{180};
\filldraw [black] (v1) circle (1.5pt)
                (v2) circle (1.5pt);
\end{scope}
\begin{scope}[xshift=3.5cm]
\coordinate[] (v1);
\coordinate[left=of v1] (a1);
\coordinate[right=1.0cm of v1] (v2);
\coordinate[right=of v2] (b1);
\draw[fermion] (a1)node[left]{$\psi$} -- (v1);
\draw[fermion](v2)--(b1)node[right]{$\bar{\psi}$};
\draw[fermion](v1)--(v2);
\semiloop[heavyboson]{v1}{v2}{10};
\filldraw [black] (v1) circle (1.5pt)
                (v2) circle (1.5pt);
\end{scope}
\begin{scope}[xshift=6.5cm]
\coordinate[] (v3);
\coordinate[below left=of v3] (v1);
\coordinate[below right=of v3] (v2);
\coordinate[position=-120 degrees from v1] (a1);
\coordinate[position=-60 degrees from v2] (a2);
\coordinate[above= of v3] (b1);
\draw[fermion](a1)node[below]{$\psi$}--(v1);
\draw[fermion](v2)--(a2)node[below]{$\bar{\psi}$};
\draw[heavyboson] (v1)-- (v2);
\draw[fermion](v1)--(v3);
\draw[fermion](v3)--(v2);
\draw[heavyboson](v3)-- (b1) node[above]{$\mathcal{V}$};
\filldraw (v1) circle (1.5pt);
\filldraw (v2) circle (1.5pt);
\end{scope}
\end{tikzpicture}
\caption{Diagrams relevant for the renormalization of $g_\calV$.}
\label{fig: vertexrun}
\end{figure}
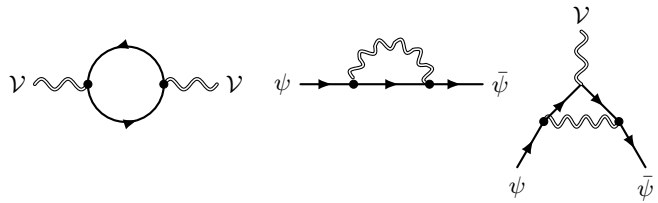

\vspace{0.25cm}
{\bf (1) Landau poles and triviality.} The $\beta$ function of the coupling of the vector fields to the fermion bi-linears (denoted $g_{\calV}$ in Eqn.~(\ref{sampleV}))
is determined by renormalizing the fermion fields and vector field two point functions, and subsequently extracting
the $\beta$ function for $g_{\calV}$. We relate the bare $(0)$ and renormalized $(r)$
fields and couplings as
\bea
V_{\mu}^{(0)} &=& \sqrt{Z_V} \, V_{\mu}^{(r)}, \quad g_{\calV}^{(0)} = Z_{g_\calV} \, g_{\calV}^{(r)} \, \mu^\eps, \\
\psi_i^{(0)} &=& \sqrt{Z_{\psi_i}} \, \psi_i^{(r)},
\eea
where $Z_x= 1 +  \delta_{Z_x}$ for $x = \{V, g_\calV, \bar{\psi}, \psi\}$. We use a renormalization scheme employing $\rm \overline{MS}$
subtraction and $d = 4 - 2 \epsilon$ dimensions using standard methods.
The relevant diagrams are shown in Figure~\ref{fig: vertexrun}.
The $\beta$-function for the running of the coupling $g_\calV$ is given by
\beq
\label{betafunceqn}
\beta_{g_\calV} = 2 \, g_\calV \,  \epsilon \, \left(-\frac{\delta_{Z_3}}{\langle \mathcal{O} \rangle} - {1 \over 2}\delta_{Z_{\bar{\psi}}} -{1 \over 2}\delta_{Z_\psi}- {1 \over 2}\delta_{Z_V}\right),
\eeq
where the renormalization factors $\delta_{Z}$'s for the various vector field cases are presented in Table~\ref{int-vectorRGE}.
The general expectation is that $g_\calV$ will have a positive $\beta$ function -- indicating Landau poles \cite{landau}, quantum triviality \cite{Wilson:1974mb} and a UV incompletion. This is indeed the case for all vector fields inducing one $(J_\psi^\mu)^\dagger \, J_{\psi,\mu}$ operator, with the exception of color $\bf \bar{3}$ vectors coupling to quark bi-linears; i.e. cases $\calV^{\bar{3}}_{\rm X,XI}$.
In this exceptional case, the $\rm SU(3)_C$ vector-fermion coupling mimics the effect of a non-abelian interaction.

An oversimplified UV scenario afflicted with an internal inconsistency
indicated by the presence of Landau poles cannot formally generate a consistent IR limit.
This indicates that further new physics must be present below the Landau pole scale $\Lambda_L$ approximated by
\bea
\Lambda_L \sim M_{\calV} \, {\rm exp} \left[g_\calV/ \beta_{g_\calV}\right].
\eea
However,
numerically corrections suppressed by $\Lambda_L$ are smaller than one loop matching effects.

\begin{figure}[t!]
\centering
\begin{tikzpicture}[node distance=0.7cm, rounded corners=0pt, line cap=round]
\begin{scope}[]
\coordinate[] (v1);
\coordinate[above left=of v1] (a1);
\coordinate[below left=of v1] (a2);
\coordinate[above right=of v1] (b1);
\coordinate[below right=of v1] (b2);
\draw[heavyboson] (a1)node[left=0.3mm]{$\mathcal{V}^{A,a}_{\mu i}$} -- (v1);
\draw[heavyboson] (a2)node[left]{$\mathcal{V}^{B,b}_{\nu j}$} -- (v1);
\draw[heavyboson](v1)--(b1)node[right]{$\mathcal{V}^{C,c}_{\rho,k}$};
\draw[heavyboson](v1)--(b2)node[right]{$\mathcal{V}^{D,d}_{\sigma,l}$};
\filldraw [black] (v1) circle (1.5pt);
\end{scope}
\begin{scope}[xshift=2.8cm]
\coordinate[] (v1);
\coordinate[above left=of v1] (a1);
\coordinate[below left=of v1] (a2);
\coordinate[right=1.0cm of v1] (v2);
\coordinate[above right=of v2] (b1);
\coordinate[below right=of v2] (b2);
\draw[heavyboson] (a1)node[left]{$\mathcal{V}$} -- (v1);
\draw[heavyboson] (a2)node[left]{$\mathcal{V}$} -- (v1);
\draw[heavyboson](v1)--(v2);
\draw[heavyboson](v2)--(b1)node[right]{$\mathcal{V}$};
\draw[heavyboson](v2)--(b2)node[right]{$\mathcal{V}$};
\filldraw [black] (v1) circle (1.5pt)
                (v2) circle (1.5pt);
\end{scope}
\begin{scope}[xshift=6.5cm,yshift=0.5cm]
\coordinate[] (v1);
\coordinate[left=of v1] (a1);
\coordinate[below=1.0cm of v1] (v2);
\coordinate[left=of v2] (a2);
\coordinate[right=of v1] (b1);
\coordinate[right=of v2] (b2);
\draw[heavyboson] (a1)node[left]{$\mathcal{V}$} -- (v1);
\draw[heavyboson] (a2)node[left]{$\mathcal{V}$} -- (v2);
\draw[heavyboson](v1)--(v2);
\draw[heavyboson](v1)--(b1)node[right]{$\mathcal{V}$};
\draw[heavyboson](v2)--(b2)node[right]{$\mathcal{V}$};
\filldraw [black] (v1) circle (1.5pt)
                (v2) circle (1.5pt);
\end{scope}
\end{tikzpicture}
\caption{$2\rightarrow2$ vector scattering diagrams.}
\label{fig: vectorscattering}
\end{figure}
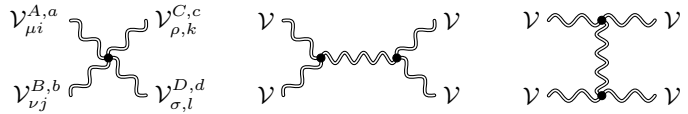

\vspace{0.25cm}
{\bf (2) Unitarity and vector self-interactions.}
 A more intractable problem is generated by $\mathcal{O}(1)$ self interactions of orphan vector fields. The four point vector self interaction is not forbidden by any symmetry. Conversely the three point interaction can be forbidden by the presence of a $\rm U(1)_Y$ charge in the composite field. Consider the $2\to2$ longitudinal vector scattering displayed in Fig.~\ref{fig: vectorscattering} that is dictated by such a four-point and three-point interactions at tree level. The relevant Lagrangian involving a general vector field with self-interactions is
\beq
\Delta \mathcal{L}_{\rm V} = {\lambda \over 4} \calV_\mu^\dagger \calV^\mu \calV_\nu^\dagger \calV^\nu + g' \,  \partial_\mu \calV^\mu \calV_\nu^\dagger \calV^\nu +\dots
\eeq
The amplitudes at leading order with the high-energy approximation for the vector polarization $\epsilon_L^\mu \simeq p^\mu/M_\calV$ through  a s-, t- and u- channel vector exchange and a four-point contact interaction, respectively, read
\bea
\mathcal{M}^L_{3,\rm s} &=& (g')^2 F_s {st-su\over 4 \, M_\calV^4},\\
\mathcal{M}^L_{3,\rm t} &=& (g')^2 F_t {st-ut\over 4 \, M_\calV^4},\\
\mathcal{M}^L_{3,\rm u} &=& (g')^2 F_u {us-ut\over 4 \, M_\calV^4},\\
\mathcal{M}^L_{4} &=&\lambda \left( F_s{t^2-u^2 \over 4 \, M_\calV^4} + F_t {s^2-u^2 \over 4 \, M_\calV^4} + F_u {s^2-t^2 \over 4 \, M_\calV^4} \right).
\eea
Here abstract group structure constants $F_{s,t,u}$ for three channels have been introduced. For example, in the model $\calV^{(1)}_{\rm IX}$: $F_s= f^{ABE} f^{CDE} f^{ijn} f^{kln}$ where ${A,B,C,D,E}$ refer to the flavor index and ${i,j,k,l,n}$ denote the iso-spin index.

If $\lambda=(g')^2$ is accomplished by a global symmetry then the amplitudes $\mathcal{M}^L_{3}$ will cancel with three terms in $\mathcal{M}^L_{4}$ with an identical $F$ factor respectively, through the Mandelstam relation $s+t+u = 4 \, M^2_\calV$. As a result, the leading scaling in $\sim (p^2)^2 /M_\calV^4$ disappears. The full amplitudes then grow as $\sim p^2/M_\calV^2$. However, if the three-point interaction is forbidden - for example due to the field carrying
a $\rm U(1)_Y$ charge - then the amplitude cannot be so moderated in its growth at high energies, and scales as $\sim (p^2)^2/M_\calV^4$.
In this manner, the presence of a $\rm U(1)_Y$ charge forbidding the scalar current simultaneously
turns off the three-point interaction that is required to moderate the high energy scattering behavior of an orphaned vector field.

Standard partial wave unitarity arguments \cite{Lee:1977eg,Schuessler:2007av,Hedri:2014mua} give
that the unitarity violation scale associated with the vector field without a three point interaction is
\bea\label{unitarity}
\Lambda \lesssim 0.2 \, M_\calV (F_t + F_u)^{-1/4} \lambda^{-1/4},
\eea
where $F_t$, an $F_u$ are determined by a particular scattering cross section.
A quick onset of unitarity violation follows from a sizable four-point interaction that is expected to emerge from a strongly interacting composite sector on general grounds.
Even introducing a loop suppression to the vector self interaction, that is $\lambda \sim (16 \pi)^{-1}$, is of little help - one still finds $\Lambda \sim M_\calV$ due to the presence of a fourth root in Eqn.~(\ref{unitarity}).
Hence, the UV strong sector should be simultaneously considered to define a consistent matching onto the SMEFT.
This would increase the low energy operator profile of such a scenario in the SMEFT beyond one operator generically due to non perturbative matchings,
and a ``one at a time" analysis invoking a tree level matching would be logically incoherent.

\vspace{0.25cm}
{\bf (3) Siblings of massive vectors with non-trivial representations.}
A massive vector field with non-trivial representations under subgroups of $\rm G$
is also generically accompanied by more ``sibling" fields.
If the massive vector gains a mass by a UV Higgs mechanism, the corresponding sibling field includes at least a scalar ($S$) obtaining a vacuum expectation value (vev).
Define this expectation value as $\langle S^\dagger S \rangle = v'^2/2$. We require $\rm  dim(\calV)  + 1 \leq dim(S)$ so that all of the components of the vector
become massive in the presence of a scalar field obtaining a vev, through eaten Goldstone components of $S$.\footnote{An additional non goldstone incomplete scalar multiplet is famously required when introducing a vev in this manner \cite{Higgs:1964pj,Englert:1964et,Guralnik:1964eu}.}

One can use the global symmetry rotations on $S$ to rotate the new vev to a (uneaten) component of $S$, denoted $s$.
The interaction of $s$ with $H^\dagger H$ cannot be forbidden
by an explicit $\rm G$ breaking without violating our assumptions. This would introduce highly constrained low energy effects into the SMEFT
through the vev $v'$ leading to the vector mass matrix.
A vacuum misalignment \cite{Kaplan:1983fs} is assumed to make the vector mass matrix symmetric under $\rm G$ in this work.
This results in the Higgs portal coupling not being suppressed by a $\rm G_Q \times G_L$ breaking spurion.
Concretely consider the Lagrangian
\bea
\mathcal{L}_{SH} = (D^\mu S)^\dagger \, (D_\mu S) - \frac{\lambda'}{4}(S^\dagger S - \frac{v'^2}{2})^2 + \lambda_{SH} S^\dagger \, S \, H^\dagger \, H. \nonumber \\
\eea
Here the covariant derivative is $D^\mu = \partial^\mu + i g_\calV \calV_a h^a$ with $h_a$ an abstract group generator that defines the non-trivial representations
that the $\calV$ multiplet carries. $S$ is expanded as $S = \left(\cdots, v' + s + \cdots\right)/\sqrt{2} + h'_a \, \rho_a$ where $\rho_a$ corresponds to the goldstone components
of the $S$ multiplet that are eaten to generate the vector mass, and the $\cdots$ fill out the full dimension of $S$. The vev $v'$ must be arranged to break the $\rm dim(\calV)$ $h'_a$ generators. Simultaneously $v'$ must not break the $\rm G$ subgroup, so $g_a \langle S \rangle = 0$,
where the generators of $\rm G$ are denoted $g_a$.\footnote{In general one expects the symmetry breaking pattern to be such that there will be uneaten goldstone bosons,
or additional massive vectors in the spectrum. Here we are considering an exceptional minimal spectrum when examining the one operator question.} Integrating out $s$ after UV symmetry breaking gives
\bea
\label{massivehiggs}
\Delta \mathcal{L}_6 =  - \frac{2 \, \lambda_{SH}^2}{\lambda' m_s^2} Q_{H \Box} - \frac{4 \, g_\calV^2}{\lambda'}(\calV_\mu \, \calV^\mu)^2 + \cdots
\eea
in addition to the operators induced by integrating out the vector field. Here the scalar mass is $m_s^2  = \lambda' \, v'^2/2$.
In addition, $\mathcal{L}_4$ terms are induced that require a finite redefinition of $\lambda$ and $v$ in the SM
to rearrange $\mathcal{L}_{\rm SM}$ back into standard from. Here we have neglected many higher order effects
including subdominant mass splitting terms. Note the sizable vector four point interaction, that is enhanced in the $\lambda' \rightarrow 0$
limit, indicating unitarity violation when the UV Higgs is integrated out of the spectrum. In this limit it is of interest to not neglect
mass splitting effects proportional to $\lambda_{SH}$.

In order to avoid assuming a UV Higgs mechanism, we can consider a composite massive vector generated by a hypothetical UV strong sector, with spin-1/2 constituents $\Psi$, so that the vector fields are $\calV^\mu \sim \langle \bar{\Psi} \gamma^\mu \, \Psi \rangle$ condensates. This composite field carries at least one non-trivial representation under one of the groups $\rm G_Q,G_L, SU(3)_C$ or $\rm SU(2)_L$ to reduce the SMEFT operator profile to one operator. Denote this non-trivial representation as $\rm \bf N$, and the corresponding group as $\rm G'$. The $\Psi$ are charged under $\rm G'$ or a larger group $\rm H$
with $\rm H \supset G'$.

We can consider $\rm G'$ or the proper subgroup case where $\rm G' \subset H$ without loss of generality with the following arguments.
The $\Psi$ belongs to $\rm SU(3)$, and ${\bf N} \in  \{{\bf 3, \bar{3}, 6,8}\}$, or $\rm SU(2)$ with ${\bf N} = \{{\bf 2,3}\}$ for the vector fields of interest. The non-trivial representations in ${\bf N}$ can be generated
from tensor products of the $\Psi$ irreducible representations. In the case where the $\Psi$ belongs to $\rm SU(3)$ we denote the irreducible representations as $\bf P, R$, which need not have the same dimension.
When ${\bf N}$ is generated by ${\bf P \otimes \bar{P}}$ the singlet representation is also generated in the tensor product. A color singlet sibling under $\rm SU(3)_C$ is expected with a mass proximate to a color octet vector, which induces a number of operators in $\mathcal{L}_6$ when integrated out.
Similarly, a flavor singlet sibling under a flavor group is also expected for flavor octets.
Interestingly, the flavor $\bf 8$ vector fields we considered all have zero $\rm U(1)_Y$ charges, so their flavor singlet siblings with the same $\rm U(1)_Y$ charge are not forbidden by the flavor symmetry to have the coupling with the corresponding quark bi-linear and also with the $J_{H, \mu}$ of vanishing $\rm U(1)_Y$ charge, inducing more than one operator in $\mathcal{L}_6$ when integrated out.
When ${\rm \bf N } \in {\bf P \otimes P}$ multiple representations result, for example in the case of $\bf P = 3$, the $\bf \bar{3}$ and $\bf 6$ fields are simultaneously present. Such fields ($\calV_{\rm X,XI}$) can induce the same operator when integrated out.
On the other hand, {\it these fields necessarily carry $\rm U(1)_Y$, and thus have a cut off scale proximate to the massive vectors mass scale for the cases consistent with our assumptions.}\footnote{This is a generic expectation if the $\Psi$ are charged under $\rm U(1)_Y$.}
Next we consider the cases when the non-trivial representation is generated by bi-linears of $\Psi$ carrying representations of unequal dimension
${\bf N} \in {\bf P \otimes R}$. By inspection of the tensor products of $\rm SU(3)$ with triality $0$ and $1$ \cite{Slansky:1981yr} it is possible to generate each ${\rm \bf N} \in  \{{\bf 3, \bar{3}, 6,8}\}$ for $\rm SU(3)$
in such a manner. However, for each $\bf P$ and $\bf R$ one can also form a condensate $\langle  \bar{\Psi} \gamma_\mu \Psi \rangle$ with zero $\rm U(1)_Y$ charge from the product ${\bf P \otimes \bar{P}}$, ${\bf R \otimes \bar{R}}$. Two more pure singlet spin one bound states proximate in mass to $M_\calV$ are expected in the spectrum, unless forbidden by another symmetry.\footnote{For example, some of the expected spectra degeneracy can be lifted in analogy to the $\eta'$ \cite{'tHooft:1976up}.}
Restricting the discussion for non-trivial $\rm SU(2)$ representations to the vector cases that do not carry $\rm U(1)_Y$ and induce one operator
at tree level, we are left with the field~$\calV_{\rm IX}^{1}$. Further, $\calV_{\rm IX}^1$ has a large flavor breaking spurion proportional to the top Yukawa generating more operators at tree level when integrated out, see Table \ref{vectors1}.

For all of these reasons, orphaned vector fields with non-trivial representations of the SM symmetry groups demand siblings and a ``good UV home".

\subsection{B. Spin $1/2$ states}

If heavy spin-1/2 states are integrated out, the mass\footnote{Here we are referring to the dominant component of the mass of the fermion
from the new sector. In addition, there will be mass contributions and splitting proportional to $\sim v$. As such effects do not act to reduce the IR SMEFT operator profile, we neglect these contributions.} of the massive fermion(s) (denoted $M$ with $M \gg v$)
must be introduced in some manner.
As discussed in the previous section, a chiral fermion with a UV Higgs mechanism induces more operators at tree level when integrating out the UV scalar field.
In this section, we confine the discussion to general vector like fermions. The general Lagrangian associated with a pair of heavy vector-like quark (VLQ) denoted by $\calQ_L, \calQ_R$ that are flavor singlets includes
\beq\label{vecquark}
\calL_{\calQ}= \calL_\calQ^0 + \calL_\calQ^\mathrm{int},
\eeq
where
\beq
\calL_\calQ^0=\bar{\calQ}_L i\sl{D} \calQ_L + \bar{\calQ}_R i \sl{D} \calQ_R - M \left( \bar{\calQ}_L \calQ_R + \bar{\calQ}_R \calQ_L \right)
\eeq
for the $\rm SU(2)_L$ singlet and doublets and
\beq
\calL_\calQ^0=\text{Tr}\left[ \bar{\calQ}_L i\sl{D} \calQ_L + \bar{\calQ}_R i \sl{D} \calQ_R\right] - M \text{Tr}\left[ \bar{\calQ}_L \calQ_R + \bar{\calQ}_R \calQ_L\right]
\eeq
for the $\rm \bf 3$ of  $\rm SU(2)_L$. The interaction term $\calL_\calQ^\mathrm{int}$ for the
VLQs to the SM fermions through the Higgs doublet is defined
as
\beq
\calL_\calQ^\mathrm{int} = J^\calQ_{L} \, \calQ_R  + J^\calQ_{R} \calQ_L   + {\rm h.c.}
\label{LagQ1}
\eeq
The requirement that the action be stationary under variations of the heavy VLQ fields $\bar{\calQ}_L, \bar{\calQ}_R$ results in two coupled EOMs:
\bea
i \sl{D} \calQ_L-M \calQ_R + (J^\calQ_R \, \gamma^0)^\dagger &=& 0, \label{eomQl} \\
i \sl{D} \calQ_R-M\calQ_L + (J^\calQ_L  \gamma^0)^\dagger &=& 0. \label{eomQr}
\eea
Mathematically, the coupled Eqns.~(\ref{eomQl}) and (\ref{eomQr}) can be solved iteratively. Taking the limit of large $M$, one can expand the classical solutions schematically as
\bea
\calQ_R &=&  {(J^\calQ_R \, \gamma^0)^\dagger \over M} + {i \sl{D}\over M^2} (J^\calQ_L \, \gamma^0)^\dagger  + \cdots, \label{QRsol}\\
\calQ_L &=& {(J^\calQ_L \, \gamma^0)^\dagger \over M} + {i \sl{D}\over M^2} (J^\calQ_R \, \gamma^0)^\dagger  + \cdots. \label{QLsol}
\eea
When substituted back into Eqn.~(\ref{vecquark}) the effect of the leading term in these solutions vanishes due to chirality.
\begin{table}[t]
\hspace*{-5mm}
\begin{center}
\begin{tabular}[t]{|c|c|c|c|c|c|c|c|c|c|}
  \hline
  Case & $\rm SU(2)_{L}$ & $\rm U(1)_Y$  & $J^\calQ_L$ & $\small Q_{u H}$ & $ \small Q_{d H}$ & $\small Q^{(1)}_{H q}$ & $ \small Q^{(3)}_{H q}$ & \\
  \hline
$\calQ_{\rm I}^{(1)}$  & 1& $-{1 \over 3}$  & $\bar{Q}_L H$ & & $\surd$ & $\surd$ & $\surd$ &   \\[2pt]
$\calQ_{\rm I}^{(3)}$  & 3& $-{1 \over 3}$  & $\sigma^{I} \bar{Q}_L H$ & $\surd$ & $\surd$ &$\surd$ &$\surd$ &  \\[2pt]
$\calQ_{\rm II}^{(1)}$  & 1 & ${2 \over 3}$   &  $\bar{Q}_L H^*$ &$\surd$ & & $\surd$&$\surd$ &   \\[2pt]
$\calQ_{\rm II}^{(3)}$  & 3 & ${2 \over 3}$   &  $\sigma^{I} \bar{Q}_L H^*$ & $\surd$& $\surd$& $\surd$& $\surd$ & \\[2pt]
  \hline
  \hline
    Case  & $\rm SU(2)_{L}$ & $\rm U(1)_Y$  & $J^\calQ_R$ & $\small Q_{u H}$ & $ \small Q_{d H}$ & $\small Q_{H u}$ & $\small Q_{H d}$ & $\small Q_{H ud}$ \\
  \hline
$\calQ_{\rm III}$  & 2& ${1 \over 6}$  &  $\bar{u}_R {H}^T$  & $\surd$&$\surd$ &$\surd$ &$\surd$ &$\surd$\\[2pt]
$\calQ_{\rm IV}$  & 2& ${1 \over 6}$  & $\bar{d}_R H^\dagger $  & $\surd$&$\surd$ &$\surd$ &$\surd$ &$\surd$\\[2pt]
$\calQ_{\rm V}$  & 2& ${7 \over 6}$  &  $\bar{u}_R H^\dagger $  &$\surd$ & & $\surd$& &\\[2pt]
$\calQ_{\rm VI}$  & 2& $-{5 \over 6}$  &  $\bar{d}_R H^T$ & & $\surd$ & &$\surd$ &\\[2pt]
  \hline
  \end{tabular}
\end{center}
\caption{\label{quarks1} Tree level $\mathcal{L}_6$ operators induced in the SMEFT with massive quarks integrated out.}
\end{table}
\begin{table}[t]
\hspace*{-5mm}
\begin{tabular}[t]{|c|c|c|c|c|c|c|c|c|c|}
  \hline
   Case  & $\rm SU(2)_{L}$ &$\rm U(1)_Y$  & $\rm G_Q$  & $J^\calQ_R$ & $Q_{u H}$ & $Q_{d H}$ & $Q_{H u}$ & $Q_{H d}$ \\
  \hline
$\calQ_{\rm VII}$  & 2& ${1 \over 6}$  &(3,1,1) &  $\bar{u}_R H^T$ & $\surd$& &$\surd$ & \\[2pt]
$\calQ_{\rm VIII}$  & 2& ${1 \over 6}$  &(1,3,1) & $\bar{d}_R H^\dagger $ &  &$\surd$ &&$\surd$ \\[2pt]
  \hline
  \end{tabular}
\caption{\label{quarks2} Tree level $\mathcal{L}_6$ operators induced in the SMEFT with massive quarks integrated out
in some sample cases with flavor quantum numbers, see Refs.~\cite{Grossman:2007bd,Gross:2010ce,Arnold:2010vs} for more discussion on the phenomenology of these fields.}
\end{table}
\begin{table}[h!]
\begin{center}
\begin{tabular}{|c|c|c|c|c|c|c|c|}
  \hline
   Case  & $\rm SU(2)_{L}$ &$\rm U(1)_Y$  & $J^\calL_L$& $Q^{(1)}_{H l}$ & $Q^{(3)}_{H l}$ & $Q_{e H}$ & $Q^{(1)}_{H e}$  \\
  \hline
$\calL_{\rm I}^{(1)}$  & 1& $-1$  & $\bar{L}_L \, H$ &$\surd$ & $\surd$ & $\surd$ &   \\[2pt]
$\calL_{\rm I}^{(3)}$  & 3& $-1$  & $\sigma^I \bar{L}_L H$ & $\surd$ & $\surd$ &$\surd$ &\\[2pt]
  \hline
  \hline
   Case  & $\rm SU(2)_{L}$ &$\rm U(1)_Y$  & $J^\calL_R$& $Q^{(1)}_{H l}$ & $Q^{(3)}_{H l}$ & $Q_{e H}$ & $Q^{(1)}_{H e}$  \\
  \hline
$\calL_{\rm III}$  & 2& $-{1 \over 2}$  &  $\bar{e}_R H^\dagger $ & & &$\surd$ & $\surd$  \\[2pt]
$\calL_{\rm IV}$  & 2& $-{3 \over 2}$  &  $\bar{e}_R H^T$ & & &$\surd$  & $\surd$ \\[2pt]
  \hline
  \end{tabular}
\end{center}
\caption{\label{leptons}  Tree level $\mathcal{L}_6$ operators induced in the SMEFT with massive leptons integrated out.}
\end{table}

We generically find that multiple operators are induced at tree level when integrating out a vector like
fermion. The cases where the vector like quark do not carry flavor quantum numbers are shown in Table~\ref{quarks1}.
In the cases that the VLQs carry flavor quantum numbers, previously discussed in Refs.~\cite{Grossman:2007bd,Gross:2010ce,Arnold:2010vs},
multiple operators are again obtained. We show some sample cases of this type in Table~\ref{quarks2}. Multiple operators at tree level are also obtained in the case of integrating out vector like leptons, see Table~\ref{leptons}.

\subsection{C. Spin 0 states}

Unlike the cases of massive vectors and spin-1/2 fields, a massive scalar can couple into the SM
through a number of interactions and naively generate many operators in the IR SMEFT matching limit.
However, the examples ($\calS_{A}$, $\calS_{B}$ and $\calS_{C}$ in Table \ref{scalars1}) discussed in Refs.~\cite{Henning:2014wua,Gorbahn:2015gxa}
show that only one operator $Q_H$, can be obtained if an explicit scale is introduced without a dynamical origin to give the
scalar a mass. For instance, $\calS_{A}$ couples through linear and bilinear interactions in the full multi-scalar potential, denoted $V(H,\calS_{A})$ in Table \ref{scalars1}.
To reduce the operator profile of $\calS_{A}$ to one operator, it is assumed that $\calS_{A}$ has a discrete or additional $\rm U(1)$ symmetry.
Such a symmetry forbids a large number of four-fermion operators at tree level, and also a number of linear $\calS$ interactions in the scalar potential that otherwise generate $Q_{H \Box}$.
Similarly,  $\calS_{B,C}$ also have minimal one operator profiles containing
only $Q_H$. However, this again follows from the UV scale being introduced without a dynamical origin.
In all these cases, a hierarchy problem in the UV sector is also introduced.

Table \ref{scalars1} also lists the cases of flavor singlet scalar fields that couple to through the $\calS^2 H^\dagger H$ interaction and in addition have an independent $\calS H^\dagger H$ interaction via a dimensionfull coupling. In these cases, the
operators $Q_{H}$ and $Q_{H \Box}$ are simultaneously produced in tree level matchings.

As in the case of massive vectors and fermions, scalars can carry non-trivial representations under $\rm G_Q$ or $\rm G_L$ to isolate the coupling to a single fermion bi-linear. These states have been studied previously in Refs.~\cite{Buchmuller:1986zs,Davies:1990sc,Arnold:2009ay,Arnold:2013cva,Arnold:2012sd}.
To avoid an explicit breaking of $\rm G_Q$ or $\rm G_L$ in this coupling,
all of these states carry at least two non-trivial representations under the flavor ($\rm G_Q$ or $\rm G_L$) or gauge ($\rm SU(3)_C$ or $\rm SU(2)_L$) groups.
For instance, consider integrating out ``di-quark" states of this form discussed in Ref.~\cite{Arnold:2009ay} at tree level.
A scalar current operator of the form $\bar{\psi}_{1L} \psi_{2R} \bar{\psi}_{2R} \psi_{1L}$ is directly obtained. This operator can be projected into the Warsaw basis via Fierz transformation,
\beq
\label{eq:genfierz}
(\bar {\psi}_{1L}  \psi_{4R}) (\bar {\psi}_{3R}  \psi_{2L}) =-{1 \over 2} (\bar {\psi}_{1L} \gamma_\mu  \psi_{2L}) (\bar {\psi}_{3R}  \gamma^\mu \psi_{4R}).
\eeq
As the ``di-quark" scalars are in non-trivial representations under $\rm SU(2)_L$ and/or $\rm SU(3)_C$ groups, the index associated with these symmetries are not contracted between the fermions in each vector current, c.f. the right hand side of Eqn.~(\ref{eq:genfierz}).
When reducing to the Warsaw basis one uses the $\rm SU(3)$ and $\rm SU(2)$ relations
\bea
T^A_{ij} T^A_{kl} &=& {1\over 2} \delta_{il} \delta_{jk} - {1\over 6} \delta_{ij} \delta_{kl}, \\
\sigma^I_{jk} \, \sigma^I_{mn} &=& 2 \delta_{jn} \, \delta_{mk} - \delta_{jk} \, \delta_{mn}.
\eea
Concretely, performing this mapping for the ``di-quark" scalars that couple to $\bar{u}_R Q_L$ and $\bar{d}_R Q_L$ induce the operators $Q_{qu}^{(1,8)}$ and $Q_{qd}^{(1,8)}$ respectively.
Similarly, the ``di-quark" scalars coupling to $Q_L Q_L$ generate $Q_{qq}^{(1,3)}$.
On the other hand, exceptional cases that can generate only one operator do exist in ``di-quark" scalars that couple to right handed $\rm SU(2)_L$ bi-linears of the same fermion field i.e. to pairs of $u_R$, $d_R$ and $e_R$. These scalars can induce the single operator $Q_{uu}$, $Q_{ee}$, $Q_{dd}$ that are defined in the Warsaw basis, see the examples in Table~\ref{scalars2}.

\begin{table}[t!]
\begin{center}
\begin{tabular}{|c|c|c|c|c|c|}
  \hline
   Case  &  $\rm SU(2)_{L}$ &$\rm U(1)_Y$  & Couplings & $Q_{H}$ & $Q_{H \Box}$   \\
  \hline
$\calS_{A}$  & 2& $1/2$  & $V(H,\calS_A)$ & $\surd$ &  \\[2pt]
$\calS_{B}$  & 4& $3/2$  & $ (H^3)^\dagger \, \calS_B+ \rm h.c.$ & $\surd$ &  \\[2pt]
$\calS_{C}$  & 4& $1/2$  & $H^\dagger \calS_C H^\dagger H  + \rm h.c.$ & $\surd$ &  \\[2pt]
$\calS_{\rm I}^{1}$  & 1& $0$  & $(\Lambda_S \calS_{\rm I} + (\calS_{\rm I})^\dagger \calS_{\rm I}) H^\dagger H$ &$\surd$ & $\surd$    \\[2pt]
$\calS_{\rm I}^{3}$  & 3& $0$  & $\Lambda_S \calS_{\rm I} \sigma H^\dagger H$, $(\calS_{\rm I})^\dagger \calS_{\rm I} H^\dagger H$ &$\surd$ & $\surd$    \\[2pt]
$\calS_{\rm II}^{1}$  & 1& $-1$  & $\Lambda_S \, \calS_{\rm II}\,H^T H$, $(\calS_{\rm II})^\dagger \calS_{\rm II} H^\dagger H$ & $\surd$ & $\surd$ \\[2pt]
$\calS_{\rm II}^{3}$  & 3 & $-1$  & $\Lambda_S \calS_{\rm II} \sigma H^T H$, $(\calS_{\rm II})^\dagger \calS_{\rm II} H^\dagger H$ & $\surd$ & $\surd$ \\[2pt]
  \hline
  \end{tabular}
\end{center}
\caption{\label{scalars1} $\mathcal{L}_6$ operators obtained at tree level when flavor and colour singlet scalars are integrated out. $\Lambda_S$ indicates a dimensionfull coupling.}
\end{table}
\begin{table}[t]
\begin{center}
\begin{tabular}{|c|c|c|c|c|c|c|}
  \hline
   Case &$\rm SU(3)_C$ & $\rm SU(2)_{L}$ &$\rm U(1)_Y$  & $\rm G_Q$ & Couples to & Op \\
  \hline
$\calS_{\rm III}$  &    3& 1& -4/3 & (3,1,1) & $u_R \, u_R$ & $ Q_{uu}$  \\
$\calS_{\rm IV}$ &  $\bar{6}$ &1& -4/3 & (${\bar 6}$,1,1) & $u_R \, u_R$ & $ Q_{uu}$ \\
$\calS_{\rm V}$ &    3& 1& 2/3 & (1,3,1)  & $d_R \, d_R$ & $ Q_{dd}$\\
$\calS_{\rm VI}$ &    $\bar{6}$ &1& 2/3 & (1,${\bar 6}$,1)  & $d_R \, d_R$ & $ Q_{dd}$ \\
\hline
\hline
  Case &$\rm SU(3)_C$ & $\rm SU(2)_{L}$ &$\rm U(1)_Y$  & $\rm G_L$ & Couples to & Op \\
\hline
$\calS_{\rm VII}$ &   1 & 1 & 2 & (1,${\bar 6}$)  & $e_R \, e_R$ & $ Q_{ee}$ \\
  \hline
\end{tabular}
\end{center}
\caption{The cases where a single $\mathcal{L}_6$ operator is generated at tree level for different scalar representations that are not singlets under the flavor group,
without the insertion of spurion Yukawa fields, from Ref.~\cite{Arnold:2009ay}.}
\label{scalars2}
\end{table}

In spite of only $Q_H$ being induced at tree level in cases $\calS_{A,B,C}$ and only one of the operators $Q_{uu}$, $Q_{dd}$ and $Q_{ee}$ obtained at tree level in the cases $\calS_{\rm III}$ - $\calS_{\rm VII}$, the arguments based on the mass scale generation from a UV Higgs mechanism with an associated extra scalar degree of freedom still hold.
The heavy scalar ($\calS$) can be embedded in a larger scalar multiplet $\calS'$ that develops a vev,
or not so embedded, when a UV Higgs mechanism is invoked to introduce a new scale $\Lambda \gg v$.
 Due to the fact that any field obtaining a vev with its self conjugate forms a singlet under {\rm G} this leads to $Q_{H \Box}$ (as shown in Eqn.~(\ref{massivehiggs})) in either case, in addition to any matchings of $\calS$ integrated out at tree level.

Alternatively, if a strong sector is present and the ``di-quark" scalar is composite,
then the arguments in favor of ``sibling" fields imply an extended spectrum that generically contains singlet composite states.
Additionally, in the presence of a confining strong sector, both spin-0 and spin-1 composite states are expected to be embedded in a spin tower \cite{Einhorn:1976uz}.
Finally, some form of dimensional transmutation can be used to generate a scale. This can take place in the context of weaker couplings using the Coleman-Weinberg (CW) mechanism~\cite{Coleman:1973jx}, or in the case with stronger couplings with a mechanism similar to the generation of $\Lambda_{\rm QCD}$.
Of course, the (weak coupling version) of CW requires multiple couplings
and generically a non-minimal UV particle spectrum.

It is also important to notice that there are scalar four-fermion current operators with the chiral structure $(\bar{L} R) (\bar{L} R)$ and $(\bar{L} R) (\bar{R} L)$ defined in the Warsaw basis. However, these operators are not constructed out of a pair of bi-linears with the same SM field content.
As a result, additional vector current operators are induced when the $(\bar{L} R) (\bar{L} R)$ and $(\bar{L} R) (\bar{R} L)$  operators
are obtained with a tree level exchange.
Nevertheless, the presence of multiple four-fermion operators induced at tree level
from the projection of some four-fermion scalar currents into the form of the Warsaw basis is clearly a more basis dependent conclusion than other arguments made in this paper.

\section{IV. Conclusions}

Driven by the question: ``Can one obtain only one dimension six operator in the SMEFT from a consistent tree level matching onto an unknown new physics sector?", in this paper we have examined the non-minimal character of the SMEFT.

We addressed this question using a ($\rm G$ and $\rm CP$) symmetry assumption to accommodate the large set of lower energy measurements that probe the symmetry breaking pattern of the SM into the $\rm TeV$ mass scale range and beyond.
We have focused on the tree level matchings capturing the consistent IR limit of a new physics sector.
Due to the extensive mixing of the operators in $\mathcal{L}_6$ under renormalization~\cite{Jenkins:2013zja,Jenkins:2013wua,Alonso:2013hga},
the SMEFT clearly has a non-minimal character once loop induced effects are considered, requiring many operators for consistent lower energy
data analysis.\footnote{Such studies can also require mapping the SMEFT to a lower energy Lagrangian,
as in studies of
$B$ decays. Using the mapping of the SMEFT to $C_9$ and $C_{10}$ as reported in \cite{Alonso:2014csa}
our results support simultaneous fits to $C_9$ and $C_{10}$. We do not find examples
where the combination of Wilson coefficients in the SMEFT at tree level naturally cancel out in these lower energy parameters.
This is largely due to the chirality of the relevant SMEFT operators. We thank a reviewer for a
suggestive inquiry on this point in the review process.}

We have uncovered some cases where only one operator (neglecting flavor indices) is naively induced at tree level.
These operators in the Warsaw basis are $(H^\dagger H)^3$, or of the four-fermion form, and come about due to a massive scalar or vector field having non-trivial representations under the symmetry groups of the SM. We have found that vector fields can carry non-zero $\rm U(1)_Y$ charge to reduce the operator profile by avoiding Higgs scalar currents being induced,
but these fields have severe unitarity problems due to the lack of a three-point vector self-interaction.
This indicates the presence of large non-perturbative matching corrections in addition to tree level matching effects.
On the other hand, when the massive vector fields are not charged under $\rm U(1)_Y$, flavor symmetries
can be introduced to reduce the IR SMEFT operator profile. In this case,
a spurion symmetry breaking analysis shows scalar currents are still induced, leading to more operators at tree level.
In practice, fitting to one pure lepton or down quark four-fermion operator is not as poorly motivated as fitting to up quark
four-fermion operators due to the relative magnitudes of the flavor breaking spurions in each case.

In contrast to the vector fields, the presence of scalar fields in a UV sector do not directly cause severe unitarity problems. Integrating them out could have a relatively minimal operator profile, i.e. only $Q_{uu}$, $Q_{dd}$ or $Q_{ee}$ is induced at tree level in the cases shown in Table \ref{scalars2}, and only $Q_H$ in some cases in Table \ref{scalars1}.
However, requiring a mass generation mechanism for these fields would lead to more matching contributions to the SMEFT operators. The scenario of a UV Higgs mechanism, if present, generically induces more operators that are constructed out of the SM Higgs field at tree level.
This occurs without a suppression by a $\rm G_Q \times G_L$ symmetry breaking spurion.
When a UV Higgs mechanism is avoided by assuming compositeness and a new strong interaction, we have argued that the requirement
of non-trivial representations for the vector and scalar fields to reduce the operator profile
would indicate the presence of an extended spectrum of the composite states - including singlet fields - that couple through many SM portals.
This would lead to more operators with tree level matchings when the extended spectrum is integrated out.

The SMEFT is a complicated field theory. It is natural and reasonable to seek a reduction of this complexity
to use in data analyses in the SMEFT framework. Using symmetry assumptions is widely accepted.
We have examined in this work if an alternative  ad-hoc approach of using ``one operator at a time" in data analyses
can be representative of a consistent tree level matching to an unknown new physics sector.
Our results show that the SMEFT has a non-minimal character quite generically
and thus this approach should be avoided, if possible.
To ensure the right conclusions are being drawn on
the degree of constraint on unknown UV physics sectors, multiple operators should be retained in data analyses.
Fortunately global data analyses in the SMEFT can already be performed with multiple operators, by using symmetries to simplify the number of parameters present.
Further the growing LHC data set makes such global analyses even more feasible to execute in practice.
In some cases, the resulting constraints can be relaxed by orders of magnitude~\cite{Han:2004az,Berthier:2016tkq} compared to a ``one operator at a time" analysis.
Nevertheless, our analysis shows that retaining multiple operators is preferred, and a relaxation of constraints can be required
to obtain a consistent IR limit of an underlying UV physics sector, when a dynamical origin of the UV scales introduced is demanded.

\vspace*{4mm}
\paragraph{\bf Acknowledgements}
We thank Luca Merlo for conversations. We thank Ilaria Brivio and William Shepherd for comments on the manuscript.
Y.J. and M.T. acknowledge generous support by the Villum Fonden and partial support from the Discovery center (DNRF91).
\vspace{-0.7cm}



\begin{thebibliography}{99}
\bibitem{Buchmuller:1985jz}
  W.~Buchmuller and D.~Wyler,
  Nucl.\ Phys.\ B {\bf 268}, 621 (1986).
  doi:10.1016/0550-3213(86)90262-2

\bibitem{Grzadkowski:2010es}
  B.~Grzadkowski, M.~Iskrzynski, M.~Misiak and J.~Rosiek,
  JHEP {\bf 1010}, 085 (2010)
  doi:10.1007/JHEP10(2010)085
  [arXiv:1008.4884 [hep-ph]].

\bibitem{Jenkins:2013zja}
  E.~E.~Jenkins, A.~V.~Manohar and M.~Trott,
  JHEP {\bf 1310}, 087 (2013)
  doi:10.1007/JHEP10(2013)087
  [arXiv:1308.2627 [hep-ph]].

\bibitem{Jenkins:2013wua}
  E.~E.~Jenkins, A.~V.~Manohar and M.~Trott,
  JHEP {\bf 1401}, 035 (2014)
  doi:10.1007/JHEP01(2014)035
  [arXiv:1310.4838 [hep-ph]].

\bibitem{Alonso:2013hga}
  R.~Alonso, E.~E.~Jenkins, A.~V.~Manohar and M.~Trott,
  JHEP {\bf 1404}, 159 (2014)
  doi:10.1007/JHEP04(2014)159
  [arXiv:1312.2014 [hep-ph]].

\bibitem{Lehman:2014jma}
  L.~Lehman,
  Phys.\ Rev.\ D {\bf 90}, no. 12, 125023 (2014)
  doi:10.1103/PhysRevD.90.125023
  [arXiv:1410.4193 [hep-ph]].

\bibitem{Lehman:2015coa}
  L.~Lehman and A.~Martin,
  JHEP {\bf 1602}, 081 (2016)
  doi:10.1007/JHEP02(2016)081
  [arXiv:1510.00372 [hep-ph]].

\bibitem{Henning:2014wua}
  B.~Henning, X.~Lu and H.~Murayama,
  JHEP {\bf 1601}, 023 (2016)
  doi:10.1007/JHEP01(2016)023
  [arXiv:1412.1837 [hep-ph]].

\bibitem{Henning:2015daa}
  B.~Henning, X.~Lu, T.~Melia and H.~Murayama,
  doi:10.1007/s00220-015-2518-2
  arXiv:1507.07240 [hep-th].

\bibitem{Henning:2015alf}
  B.~Henning, X.~Lu, T.~Melia and H.~Murayama,
  arXiv:1512.03433 [hep-ph].

\bibitem{Giudice:2011ak}
  G.~F.~Giudice, B.~Gripaios and R.~Sundrum,
  JHEP {\bf 1108}, 055 (2011)
  doi:10.1007/JHEP08(2011)055
  [arXiv:1105.3161 [hep-ph]].



\bibitem{Chivukula:1987py}
  R.~S.~Chivukula and H.~Georgi,
  Phys.\ Lett.\ B {\bf 188}, 99 (1987).
  doi:10.1016/0370-2693(87)90713-1

\bibitem{D'Ambrosio:2002ex}
  G.~D'Ambrosio, G.~F.~Giudice, G.~Isidori and A.~Strumia,
  Nucl.\ Phys.\ B {\bf 645}, 155 (2002)
  doi:10.1016/S0550-3213(02)00836-2
  [hep-ph/0207036].

\bibitem{Cirigliano:2005ck}
  V.~Cirigliano, B.~Grinstein, G.~Isidori and M.~B.~Wise,
  Nucl.\ Phys.\ B {\bf 728}, 121 (2005)
  doi:10.1016/j.nuclphysb.2005.08.037
  [hep-ph/0507001].

\bibitem{Jarlskog:1985ht}
  C.~Jarlskog,
  Phys.\ Rev.\ Lett.\  {\bf 55}, 1039 (1985).
  doi:10.1103/PhysRevLett.55.1039

\bibitem{Jarlskog:1985cw}
  C.~Jarlskog,
  Z.\ Phys.\ C {\bf 29}, 491 (1985).
  doi:10.1007/BF01565198


\bibitem{Weinberg:1979sa}
  S.~Weinberg,
  Phys.\ Rev.\ Lett.\  {\bf 43}, 1566 (1979).
  doi:10.1103/PhysRevLett.43.1566

\bibitem{Wilczek:1979hc}
  F.~Wilczek and A.~Zee,
  Phys.\ Rev.\ Lett.\  {\bf 43}, 1571 (1979).
  doi:10.1103/PhysRevLett.43.1571



\bibitem{Gavela:2016bzc}
  B.~M.~Gavela, E.~E.~Jenkins, A.~V.~Manohar and L.~Merlo,
  Eur.\ Phys.\ J.\ C {\bf 76}, no. 9, 485 (2016)
  doi:10.1140/epjc/s10052-016-4332-1
  [arXiv:1601.07551 [hep-ph]].

  \bibitem{Georgi:1974sy}
  H.~Georgi and S.~L.~Glashow,
  Phys.\ Rev.\ Lett.\  {\bf 32}, 438 (1974).
  doi:10.1103/PhysRevLett.32.438


\bibitem{Grinstein:2011yv}
  B.~Grinstein, A.~L.~Kagan, M.~Trott and J.~Zupan,
  Phys.\ Rev.\ Lett.\  {\bf 107}, 012002 (2011)
  doi:10.1103/PhysRevLett.107.012002
  [arXiv:1102.3374 [hep-ph]].

\bibitem{Grinstein:2011dz}
  B.~Grinstein, A.~L.~Kagan, J.~Zupan and M.~Trott,
  JHEP {\bf 1110}, 072 (2011)
  doi:10.1007/JHEP10(2011)072
  [arXiv:1108.4027 [hep-ph]].

\bibitem{delAguila:2010mx}
  F.~del Aguila, J.~de Blas and M.~Perez-Victoria,
  JHEP {\bf 1009}, 033 (2010)
  doi:10.1007/JHEP09(2010)033
  [arXiv:1005.3998 [hep-ph]].


\bibitem{landau}
L.~D.~Landau, Niels Bohr and the Development of
Physics, edited by W. Pauli (Pergamon Press, London, 1955).

\bibitem{Wilson:1974mb}
  K.~G.~Wilson,
  Rev.\ Mod.\ Phys.\  {\bf 47}, 773 (1975).
  doi:10.1103/RevModPhys.47.773

\bibitem{Lee:1977eg}
  B.~W.~Lee, C.~Quigg and H.~B.~Thacker,
  Phys.\ Rev.\ D {\bf 16}, 1519 (1977).
  doi:10.1103/PhysRevD.16.1519

\bibitem{Schuessler:2007av}
  A.~Schuessler and D.~Zeppenfeld,
  In *Karlsruhe 2007, SUSY 2007* 236-239
  [arXiv:0710.5175 [hep-ph]].

\bibitem{Hedri:2014mua}
  S.~El Hedri, W.~Shepherd and D.~G.~E.~Walker,
  arXiv:1412.5660 [hep-ph].

\bibitem{Higgs:1964pj}
  P.~W.~Higgs,
  Phys.\ Rev.\ Lett.\  {\bf 13}, 508 (1964).
  doi:10.1103/PhysRevLett.13.508

\bibitem{Englert:1964et}
  F.~Englert and R.~Brout,
  Phys.\ Rev.\ Lett.\  {\bf 13}, 321 (1964).
  doi:10.1103/PhysRevLett.13.321

\bibitem{Guralnik:1964eu}
  G.~S.~Guralnik, C.~R.~Hagen and T.~W.~B.~Kibble,
  Phys.\ Rev.\ Lett.\  {\bf 13}, 585 (1964).
  doi:10.1103/PhysRevLett.13.585

\bibitem{Kaplan:1983fs}
  D.~B.~Kaplan and H.~Georgi,
  Phys.\ Lett.\  {\bf 136B}, 183 (1984).
  doi:10.1016/0370-2693(84)91177-8

\bibitem{'tHooft:1976up}
  G.~'t Hooft,
  Phys.\ Rev.\ Lett.\  {\bf 37}, 8 (1976).
  doi:10.1103/PhysRevLett.37.8

\bibitem{Grossman:2007bd}
  Y.~Grossman, Y.~Nir, J.~Thaler, T.~Volansky and J.~Zupan,
  Phys.\ Rev.\ D {\bf 76}, 096006 (2007)
  doi:10.1103/PhysRevD.76.096006
  [arXiv:0706.1845 [hep-ph]].

\bibitem{Gross:2010ce}
  E.~Gross, D.~Grossman, Y.~Nir and O.~Vitells,
  Phys.\ Rev.\ D {\bf 81}, 055013 (2010)
  doi:10.1103/PhysRevD.81.055013
  [arXiv:1001.2883 [hep-ph]].

\bibitem{Arnold:2010vs}
  J.~M.~Arnold, B.~Fornal and M.~Trott,
  JHEP {\bf 1008}, 059 (2010)
  doi:10.1007/JHEP08(2010)059
  [arXiv:1005.2185 [hep-ph]].


\bibitem{Slansky:1981yr}
  R.~Slansky,
  Phys.\ Rept.\  {\bf 79}, 1 (1981).
  doi:10.1016/0370-1573(81)90092-2

\bibitem{Gorbahn:2015gxa}
  M.~Gorbahn, J.~M.~No and V.~Sanz,
  JHEP {\bf 1510}, 036 (2015)
  doi:10.1007/JHEP10(2015)036
  [arXiv:1502.07352 [hep-ph]].



\bibitem{Buchmuller:1986zs}
  W.~Buchmuller, R.~Ruckl and D.~Wyler,
  Phys.\ Lett.\ B {\bf 191}, 442 (1987)
  Erratum: [Phys.\ Lett.\ B {\bf 448}, 320 (1999)].
  doi:10.1016/S0370-2693(99)00014-3, 10.1016/0370-2693(87)90637-X

\bibitem{Davies:1990sc}
  A.~J.~Davies and X.~G.~He,
  Phys.\ Rev.\ D {\bf 43}, 225 (1991).
  doi:10.1103/PhysRevD.43.225

\bibitem{Arnold:2009ay}
  J.~M.~Arnold, M.~Pospelov, M.~Trott and M.~B.~Wise,
  JHEP {\bf 1001}, 073 (2010)
  doi:10.1007/JHEP01(2010)073
  [arXiv:0911.2225 [hep-ph]].

\bibitem{Arnold:2013cva}
  J.~M.~Arnold, B.~Fornal and M.~B.~Wise,
  Phys.\ Rev.\ D {\bf 88}, 035009 (2013)
  doi:10.1103/PhysRevD.88.035009
  [arXiv:1304.6119 [hep-ph]].


\bibitem{Arnold:2012sd}
  J.~M.~Arnold, B.~Fornal and M.~B.~Wise,
  Phys.\ Rev.\ D {\bf 87}, 075004 (2013)
  doi:10.1103/PhysRevD.87.075004
  [arXiv:1212.4556 [hep-ph]].

\bibitem{Coleman:1973jx}
  S.~R.~Coleman and E.~J.~Weinberg,
  Phys.\ Rev.\ D {\bf 7}, 1888 (1973).
  doi:10.1103/PhysRevD.7.1888



\bibitem{Einhorn:1976uz}
  M.~B.~Einhorn,
  Phys.\ Rev.\ D {\bf 14}, 3451 (1976).
  doi:10.1103/PhysRevD.14.3451

  \bibitem{Alonso:2014csa}
    R.~Alonso, B.~Grinstein and J.~Martin Camalich,
    Phys.\ Rev.\ Lett.\  {\bf 113} (2014) 241802
    doi:10.1103/PhysRevLett.113.241802
    [arXiv:1407.7044 [hep-ph]].

\bibitem{Han:2004az}
  Z.~Han and W.~Skiba,
  Phys.\ Rev.\ D {\bf 71}, 075009 (2005)
  doi:10.1103/PhysRevD.71.075009
  [hep-ph/0412166].

\bibitem{Berthier:2016tkq}
  L.~Berthier, M.~Bj\o rn and M.~Trott,
  JHEP {\bf 1609}, 157 (2016)
  doi:10.1007/JHEP09(2016)157
  [arXiv:1606.06693 [hep-ph]].


\end{thebibliography}
\end{document}